\documentclass[acmsmall]{acmart}

\usepackage[suppress]{color-edits} 
\usepackage{xspace}
\usepackage{color, colortbl}
\usepackage{xcolor}
\usepackage{enumitem}
\usepackage[singlelinecheck=false]{caption}
\usepackage{tabularx}  
\usepackage{multirow}
\usepackage{multicol}
\usepackage{threeparttable}
\usepackage{float}
\usepackage[bottom]{footmisc}
\newcolumntype{P}[1]{>{\centering\arraybackslash}p{#1}}
\newcolumntype{R}[1]{>{\raggedleft\let\newline\\\arraybackslash\hspace{0pt}}m{#1}}

\definecolor{dawnblue}{rgb}{0.84, 0.92, 1.0}

\newcommand{\best}[1]{\colorbox{dawnblue}{\bf #1}}

\usepackage[capitalize,noabbrev]{cleveref}

\setlist[itemize]{align=parleft,left=0.5em..1.5em}

\newlist{req}{enumerate}{2}
\setlist[req,1]{label=RQ \arabic*:,ref= \arabic*, leftmargin=*}

\newlist{hyp}{enumerate}{2}
\setlist[hyp,1]{before=\itshape,font=\itshape, label=Hypothesis \arabic*:,ref= \arabic*, leftmargin=*}
\setlist[hyp,2]{before=\itshape,font=\itshape, label=Hypothesis \arabic*:,ref= \arabic*, leftmargin=*}

\addauthor{km}{blue} 
\addauthor{ps}{blue} 
\addauthor{ap}{blue} 
\addauthor{vt}{blue} 
\addauthor{nk}{blue} 

\AtBeginDocument{%
  \providecommand\BibTeX{{%
    \normalfont B\kern-0.5em{\scshape i\kern-0.25em b}\kern-0.8em\TeX}}}

\setcopyright{rightsretained}
\acmJournal{PACMHCI}
\acmYear{2024} \acmVolume{8} \acmNumber{CSCW1} \acmArticle{183} \acmMonth{4} \acmPrice{}\acmDOI{10.1145/3641022}

\received{July 2023}
\received[revised]{October 2023}
\received[accepted]{November 2023}

\makeatletter
\gdef\@copyrightpermission{
  \begin{minipage}{0.2\columnwidth}
   \href{http://creativecommons.org/licenses/by-nc/4.0/}{\includegraphics[width=0.90\textwidth]{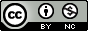}}
  \end{minipage}\hfill
  \begin{minipage}{0.8\columnwidth}
   \href{http://creativecommons.org/licenses/by-nc/4.0/}{This work is licensed under a Creative Commons Attribution-NonCommercial International 4.0 License.}
  \end{minipage}
  \vspace{5pt}
}
\makeatother

\begin{document}

\title[Impact of Imperfect XAI]{The Impact of Imperfect XAI on Human-AI Decision-Making}

\author{Katelyn Morrison}
\email{kcmorris@cs.cmu.edu}
\orcid{0000-0002-2644-4422}
\authornote{Both authors contributed equally to this research.}
\affiliation{%
  \institution{Carnegie Mellon University}
  \city{Pittsburgh}
  \state{Pennsylvania}
  \country{USA}
}

\author{Philipp Spitzer}
\email{philipp.spitzer@kit.edu}
\orcid{0000-0002-9378-0872}
\authornotemark[1]
\affiliation{%
  \institution{Karlsruhe Institute of Technology}
  \country{Germany}
}

\author{Violet Turri}
\email{vturri@andrew.cmu.edu}
\orcid{0009-0002-3081-5617}
\affiliation{%
  \institution{Carnegie Mellon University}
  \city{Pittsburgh}
  \state{Pennsylvania}
  \country{USA}
}

\author{Michelle Feng}
\email{msfeng@andrew.cmu.edu}
\orcid{0009-0002-2358-2341}
\affiliation{%
  \institution{Carnegie Mellon University}
  \city{Pittsburgh}
  \state{Pennsylvania}
  \country{USA}
}

\author{Niklas Kühl}
\email{kuehl@uni-bayreuth.de}
\orcid{0000-0001-6750-0876}
\affiliation{%
  \institution{University of Bayreuth}
  \country{Germany}
}

\author{Adam Perer}
\email{adamperer@cmu.edu}
\orcid{0000-0002-8369-3847}
\affiliation{%
  \institution{Carnegie Mellon University}
  \city{Pittsburgh}
  \state{Pennsylvania}
  \country{USA}
}



\renewcommand{\shortauthors}{Katelyn Morrison, et al.}
\begin{abstract}

Explainability techniques are rapidly being developed to improve human-AI decision-making across various cooperative work settings. Consequently, previous research has evaluated how decision-makers collaborate with imperfect AI by investigating appropriate reliance and task performance with the aim of designing more human-centered computer-supported collaborative tools. Several human-centered explainable AI (XAI) techniques have been proposed in hopes of improving decision-makers' collaboration with AI; however, these techniques are grounded in findings from previous studies that primarily focus on the impact of incorrect AI advice. Few studies acknowledge the possibility of the explanations being incorrect even if the AI advice is correct. Thus, it is crucial to understand how imperfect XAI affects human-AI decision-making. In this work, we contribute a robust, mixed-methods user study with 136 participants to evaluate how incorrect explanations influence humans' decision-making behavior in a bird species identification task, taking into account their level of expertise and an explanation's level of assertiveness. Our findings reveal the influence of imperfect XAI and humans' level of expertise on their reliance on AI and human-AI team performance. We also discuss how explanations can deceive decision-makers during human-AI collaboration. Hence, we shed light on the impacts of imperfect XAI in the field of computer-supported cooperative work and provide guidelines for designers of human-AI collaboration systems.
\end{abstract}




\maketitle

\section{Introduction}

With the deployment of imperfect artificial intelligence (AI) in high-stakes decision-making scenarios, decision-makers struggle with knowing when they should and should not rely on AI advice, causing frustration and resulting in potentially harmful decisions. As a result, designing and developing human-centered explanations has become a core theme in computer-supported cooperative work (CSCW) and human-AI collaboration research~\cite{wang2019designing, liao2021human,ehsan2022human}. Tangentially, recent work has proposed new explanation techniques that leverage machine learning models to explain the prediction of another machine learning model~\cite{bertaglia2023closing,kayser2022explaining,hendricks2021generating,tanida2023interactive,warburg2021bayesian,barata2021improving,sadeghi2018users,yang2023harnessing,tursun2023towards,kroeger2023large,sovrano2023toward}. A subset of these studies propose to exploit language models to generate natural language explanations for image classifications with the rationalization that natural language is more ``human-friendly''~\cite{hendricks2021generating,kayser2022explaining,tanida2023interactive}. Aside from natural language explanations, another subset of recent work proposes advanced content-based image recognition techniques to generate example-based explanations~\cite{warburg2021bayesian,barata2021improving,sadeghi2018users}. These types of approaches to explainability introduce another level of uncertainty in the collaboration between the decision-maker and the AI, as explanation models are imperfect.

CSCW, and more specifically human-AI collaboration, is prevalent across high-stakes scenarios~\cite{cai2019hello, wang2019human, lindvall2021rapid, tschandl2020human}. For instance, radiologists collaborate with AI to identify abnormalities in medical imagery~\cite{tschandl2020human}; conservationists use AI to help monitor biodiversity~\cite{berger2017wildbook}, and humanitarian aids use AI to help identify damaged buildings after natural disasters or armed conflicts from satellite imagery \cite{zhang2019crowdlearn, morrison-xai-23}. While some of these human-AI collaboration scenarios require the human decision-maker to have several years of experience in the given domain, such as radiology, monitoring biodiversity with the help of AI doesn't necessarily require domain expertise~\cite{berger2017wildbook,paxton2019citizen}. Platforms, such as iNaturalist~\cite{iNaturalist}, Merlin Bird App~\cite{merlin}, and Wildbooks from WildMe.org~\cite{Wildme}, have allowed non-experts (\emph{i.e.}, citizen scientists, hobbyists, or students) to partake in monitoring biodiversity alongside domain experts (\emph{i.e.}, ornithologists and conservationists). While these platforms are valuable for non-experts to use, the AI models backing these platforms are imperfect: they do not always provide correct predictions~\cite{kocielnik2019will}. 

Experts and non-experts interacting with the same imperfect AI and the same type of explanations in human-AI collaboration scenarios, such as decision-making \cite{schemmer2022meta} or learning systems \cite{spitzer2023ml}, could result in some users misunderstanding or inappropriately relying on/overriding the AI advice. Experts may have more context outside of the AI's classification and confidence that a non-expert may not have. This might result in experts using their context information to appropriately rely on the AI when advice is provided, such as correctly overriding when wrong AI advice is presented and correctly using AI advice when it is correct. Non-experts, on the other hand, might not be able to judge the correctness of the AI advice appropriately as they are missing this context information. For example, for bird species identification, ornithologists tend to be more aware of information related to the visual differences between the male and female birds for a given species, the bird's habitat, and migration patterns, whereas non-experts may not know some or all of that information. This same situation can arise in radiology where residents (``non-experts'') may initially be less familiar with certain diseases than an attending radiologist (``experts''). However, both experts and non-experts can struggle to identify certain instances. In this case, collaborating with AI can result in complementary team performance (CTP), leveraging the unique knowledge of both humans and AI, resulting in the human-AI team's task performance being better than the human or AI alone~\cite{hemmer2021human}.

Inappropriate reliance can also occur in the presence of \textit{imperfect XAI}, a term that we introduce to represent the phenomenon where an explanation reveals evidence that does not necessarily comply with the prediction. ~\citet{papenmeier2019model} use the term `explanation fidelity' while ~\citet{kroeger2023large} use the term `faithfulness' to measure how ``truthful'' an explanation is. We use imperfect XAI to align with existing terms, such as imperfect AI, in the CSCW and human-computer interaction (HCI) communities. Specifically, we define imperfect XAI as explanation techniques that can potentially provide explanations that do not fit with the AI's predictions. We view explanation fidelity or faithfulness as a term that can be used under the umbrella term of imperfect XAI; we view explanation fidelity as referring to the continuum of explanation correctness, such as when explanations are partially correct.
Imperfect XAI can exist regardless of whether the AI's advice is correct or not. AI explanations may oversimplify or improperly estimate complex models in order to make them more interpretable, deceiving and misleading the human decision-maker. As a result, non-experts may be more prone to under- or over-relying on AI advice in the presence of incorrect explanations. Within knowledge transfer scenarios, this could cause the non-expert to learn incorrect information about a given class. Furthermore, previous research shows that the language tone within natural language explanations can impact decision-makers \cite{calisto2023assertiveness}. However, the effect of language tone on humans' appropriate reliance when interacting with imperfect explanations is underexplored. Therefore, it is necessary to investigate how the communication style of explanations (\emph{e.g.}, the language tone\kmedit{, the information provided, and their representations}) impacts human-AI collaborations across levels of expertise ~\cite{kim2023communicating,calisto2023assertiveness}.

Collaborating with imperfect AI is not a new concept to the CSCW and HCI communities~\cite{hemmer2021human,lindvall2021rapid,kocielnik2019will}. Despite numerous user studies over the years investigating human-AI collaborations and XAI, few have formally acknowledged the existence of imperfect XAI in their studies~\cite{papenmeier2019model,hemmer2021human}. By formally acknowledging the existence of imperfect XAI in human-AI collaboration, research has several new interesting dimensions to explore. Although numerous user studies seek to understand how humans align, perceive, and interact with different types of explanations in various human-AI collaboration scenarios (\emph{e.g.}, ~\cite{chen2023understanding,kim2023help}), few studies explore the impact that incorrect or ``noisy'' explanations have on human-AI collaboration~\cite{vasconcelos2023generation,jang2023know,papenmeier2019model,hemmer2021human}. Recent work has used technical approaches to mitigate ``noisy'' or incorrect natural language explanations~\cite{jang2023know}, and ~\citet{kroeger2023large} propose metrics to algorithmically evaluate the effectiveness of the generated post-hoc natural language explanations. However, to our knowledge, no studies investigate how the interaction between the correctness of explanations and the decision-maker's level of expertise impact appropriate reliance on AI, human-AI team performance, and the extent to which AI explanations deceive decision-makers. 


With the growing use of machine learning models to explain other machine learning models in high-stakes decision-making scenarios, we argue that it is necessary to understand how humans interact with explanations that are incorrect, even when the AI's advice is correct. We also argue that it is important to understand the relationship that the level of expertise and the tone of explanations (\emph{i.e.}, assertive, non-assertive, or neutral) have on a decision-maker's reliance on AI. Understanding these dimensions of human-AI collaboration will provide insight to XAI and CSCW researchers. With these topics under-explored in current literature, we present the following research questions:

\begin{req}[leftmargin=1.06cm, labelindent=0pt, labelwidth=0em, label=\textbf{RQ\arabic*}:, ref=\arabic*]
    \item How does the correctness of explanations affect appropriate reliance on AI, and to what extent do the decision-maker's level of expertise and the explanation's assertiveness moderate this effect? \label{rq1}
    \item How is complementary team performance impacted by the correctness of explanations and the decision-maker's level of expertise? \label{rq2} 
    \item How do different types of explanations change the effect that the correctness of explanations has on appropriate reliance and complementary team performance? \label{rq3} 
    \item To what extent do incorrect and correct explanations deceive decision-makers with different levels of expertise? \label{rq4}
\end{req}

To address our research questions, we employ an imperfect AI model for a bird species identification task~\cite{hendricks2021generating}. We focus on bird species identification because the use of AI for wildlife conservation efforts among experts and non-experts is a rapidly growing field in research and practice~\cite{tuia2022perspectives}. \kmedit{Furthermore, it is less difficult to find people with varying levels of expertise in birding than in radiology who will have time to participate in our study.}

Through a mixed-methods study, we answer our research questions by asking participants to classify bird images in two phases: without any advice from AI (phase 1) and then showing the AI's advice and explanation (phase 2). To answer \textbf{RQ} \ref{rq1}, we design a research model based on phenomena from relevant research in CSCW and conduct rigorous moderation analyses based on \cite{hayes2017introduction}. Our analyses leverage the appropriate reliance metrics defined by ~\citet{schemmer2023appropriate}. We design our study to be within-subjects for the correctness of the explanation and the assertiveness of explanations allowing us to answer \textbf{RQ} \ref{rq2} and between-subjects for the explanation modality allowing us to answer \textbf{RQ} \ref{rq3}. Moreover, we calculate the magnitude of deception caused by incorrect explanations compared to correct explanations across both explanation modalities and levels of expertise to account for the impact of imperfect XAI on humans' decision-making behavior. This measurement gives us insight into \textbf{RQ} \ref{rq4}.
Lastly, we conduct an inductive content analysis based on \citet{gioia2013seeking} to assess the open-ended responses from participants to gain insight into designing for imperfect XAI in human-AI collaborations.

As a result of our study, we contribute the following to the CSCW community:
\begin{itemize}
    \item \textit{Research Model for Human-AI Collaboration with Imperfect XAI}: We propose a research model for the moderating roles of the decision-maker's level of expertise and the explanation's assertiveness on the effect of the correctness of explanations on appropriate reliance.
    \item \textit{Novel Empirical Study}: To validate our proposed research model, we conduct the first empirical investigation that explores the moderation of assertiveness of explanations and the level of expertise on the impact that the correctness of explanations has on appropriate reliance. We do this on a human-AI collaboration scenario across two different modalities of explanations: natural language explanations and visual, example-based explanations. We also investigate the impact on complementary team performance. Our findings inform designers of human-AI collaboration systems on how to deploy imperfect XAI from a user-centric perspective.
    \item \textit{Novel Metric for Impact of Imperfect XAI:} We contribute a novel metric to the human-AI decision-making field accounting for the impact of incorrect explanations on humans' decision-making behavior when collaborating with AI. Specifically, we propose the Deception of Reliance (DoR) caused by imperfect XAI. With DoR, we investigate to what extent imperfect XAI deceives humans.
    \item \textit{Qualitative Insights}: We provide insights on how decision-makers, regardless of expertise, prefer the tone of explanations to align with factors related to the AI's behavior and the impact on decision-makers. These insights can inform future works to conduct new evaluations.
\end{itemize}

The remainder of this article is structured as follows: We present related literature on imperfect AI systems, human-AI collaboration, and explainability (\Cref{related_section}, p. \pageref{related_section}) before outlining our theoretical development (\Cref{theoretical_section}, p. \pageref{theoretical_section}) and methodology for conducting an empirical study (\Cref{methodology}, p. \pageref{methodology}). We then present the results of our work for two different types of explanations (\Cref{results_section}, p. \pageref{results_section}). After that, we discuss the implications (\Cref{discussion_section}, p. \pageref{discussion_section}) and limitations (\Cref{limitations}, p. \pageref{limitations}) of our findings. Finally, we end our article with a brief conclusion (\Cref{conclusion_section}, p. \pageref{conclusion_section}).
\section{Related Work} 
\label{related_section}

 We situate our contributions in relation to past work about decision-making with imperfect AI/XAI, the impacts of end-user expertise on human-AI collaboration, and the impact of explanation type on human-AI collaboration. 

\subsection{Decision-Making with Imperfect AI/XAI}

Numerous studies in CSCW and HCI have investigated the impact that imperfect AI has on human-AI collaboration (\emph{e.g.}, ~\cite{kocielnik2019will, bansal2021does,vasconcelos2023generation}). 
~\citet{kocielnik2019will} offer three techniques for setting user expectations about the performance of an imperfect AI system, including an accuracy indicator, example-based explanations, and performance control. Through a user study with an AI-powered scheduling assistant, the authors demonstrate the efficacy of their techniques in maintaining user satisfaction and acceptance. The authors also demonstrate that the nature of system errors can impact user perception.

Several recent studies investigate how programmers collaborate with Copilot, an imperfect AI programming assistant (\emph{e.g.}, ~\cite{vasconcelos2023generation,barke2023grounded,dakhel2023github}). One of those studies specifically looks at how to convey the uncertainty of outputs from Copilot~\cite{vasconcelos2023generation}. By highlighting code that is most likely going to be edited by the programmer instead of highlighting based on the probability of the code being generated, they observe that programmers arrive at solutions faster. Furthermore, ~\citet{dakhel2023github} conclude that GitHub Copilot is valuable for expert programmers but something non-expert programmers should be cautious about.

Previous studies explore the impact that revealing the confidence of the model's prediction has on the human-AI team (\emph{e.g.}, \cite{kim2020effect,tejeda2023displaying, bansal2021does}). For example, \citet{kim2020effect} investigate the effect of various framings and timings for presenting the performance of an AI system on user acceptance. Through their user study, the authors reveal that users find AI advice to be more reasonable when it is not accompanied by information about AI system performance than when it is. In the case that AI system performance is shown, users consider AI advice to be more reasonable when system performance is displayed before they make a decision rather than afterward. However, communicating uncertainty for image classification in a visual format is under-explored. Recent work conducts a user study to see how showing the confidence of an AI prediction through a green hue on an image impacts reliance on AI~\cite{tejeda2023displaying}.

Fewer studies investigate the impact that imperfect XAI has on human-AI collaboration~\cite{papenmeier2019model}.
Similar to our contributions, ~\citet{papenmeier2019model} investigate the impact that explanation fidelity has on user trust. They present a user study where participants collaborate with AI of different accuracies and XAI with different levels of correctness to determine if a Tweet should be published or not based on its content. While ~\citet{papenmeier2019model} investigate how an explanation's level of correctness impacts trust, they do not explore the role that a user's level of expertise plays.

\subsection{Domain Expertise \& Human-AI Complementarity}

There has been a growing interest in understanding the impact that the decision-maker's domain expertise has on human-AI collaborations~\cite{nourani2020role,calisto2023assertiveness,szymanski2021visual,dikmen2022effects,tschandl2020human,ford2022explaining,bayer2021role,zhao2023exploring,ooge2021trust}. One recent study investigates the impact of decision-makers' domain expertise on task accuracy in a high-stakes human-AI collaboration task~\cite{calisto2023assertiveness}. ~\citet{calisto2023assertiveness} also look at the impact that the assertiveness of natural language explanations has on human-AI collaboration. In their study, they present natural language explanations with varying levels of assertiveness to radiologists with different years of experience on a mammogram classification task. Their main analysis consists of the radiologists' task performance. Unlike ~\citet{calisto2023assertiveness}, our experiment collects the human's initial decision before showing the AI's advice to the human, allowing us to measure appropriate reliance and assess for complementary team performance.

One study investigates how the level of expertise for Arabic or Indian Numerals from various versions of MNIST impacts task accuracy and model perception~\cite{ford2022explaining}. A similar study shows clinicians with various levels of expertise four different types of explanations, including visual, example-based explanations~\cite{tschandl2020human}. Similar to our study design, they show participants the three most similar example images for the example-base explanations. Another study investigates how practitioners with different levels of expertise perceive explanations that were implemented in a manufacturing industry context~\cite{zhao2023exploring}. They observe that practitioners with a higher level of expertise are more accepting of the explanations. Recent work proposes a research model to identify the impact decision-maker's level of expertise has on trust in XAI~\cite{bayer2021role}. Their research model does not consider the correctness or tone of explanations. Through their online, AI-supported chess experiment, they observe that expertise negatively affects trust. 

While numerous previous works investigate the impact of domain expertise, to the best of our knowledge, none explore the impact of the correctness of explanations and the level of expertise on the decision-maker's reliance behavior together.

\subsection{Explanation Modality}

In human-AI collaboration scenarios, the human decision-making behavior depends on the type of explanation provided (\emph{e.g.}, ~\cite{humer2022comparing}).
To validate why we evaluate our research model for two different types of explanations (\emph{i.e.}, natural language and visual, example-based), we synthesize previous work that compares multiple different modes of XAI. 


Several studies investigate the use of example-based explanations in human-AI decision-making~\cite{cai2019effects,humer2022comparing,chen2023understanding,du2022role,yang2020visual}. ~\citet{cai2019effects} propose normative and comparative explanations, different types of example-based explanations. They evaluate how these explanations impact end-users' understandability and perception of the AI model in a drawing guessing game. The authors find the normative explanations to help users better understand how the AI makes decisions when the model prediction is incorrect. Another paper investigates example-based explanations in a slightly different format from Cai~\cite{yang2020visual}. They similarly found the example-based explanations improve the users' appropriate trust in the classifier. 

In a different study, \citet{du2022role} examines the effect of different explanation modalities on clinical practitioners' reliance behavior. The authors show no significant differences between example-based explanations and feature-based explanations. However, they find that different types of practitioners prefer different modalities from a user-centric perspective. 
More recent work compares example-based explanations to feature importance through a think-aloud study~\cite{chen2023understanding}. From their mixed-methods study, ~\citet{chen2023understanding} outline three types of intuition that are employed when decision-makers reason about AI predictions and explanations, including task outcomes, features, and AI limitations. The authors use these three intuition types to explain study results in which feature-based explanations lead to overreliance on AI while example-based explanations improve human-AI performance.

Several recent works have compared text-based explanations to visual explanations (\emph{e.g.}. ~\cite{kim2023should,robbemond2022understanding,szymanski2021visual}). For example, \citet{kim2023should} analyze a unified explanation technique from a human-centric point of view. In their work, \citet{kim2023should} explore visual and text explanations in a user study. They investigate users' preferences for different AI interfaces. The authors conclude that users prefer local visual explanations in such interfaces over text-based ones. Another study compares six different types of explanation modalities in an extensive user study \cite{robbemond2022understanding}. \citet{robbemond2022understanding} look into the impact of text, audio, graphics, and combinations of the previous modalities on decision-makers' reliance on decision support systems. Their results show that combinations of different explanation modalities lead to higher user performance. ~\citet{szymanski2021visual} conduct a similar study, only evaluating visual and textual explanations. 

Based on findings from previous studies that evaluate the impact of various explanation modalities on human-AI collaboration, we choose to explore visual, example-based explanations and natural language explanations.
\section{Theoretical Development}
\label{theoretical_section}

The increasing use of explanations to reveal the rationale behind AI predictions has led to a rise in research examining the impact of explanations on decision-makers' behavior \cite{de2022perils, schemmer2022meta, leichtmann2023explainable}. As imperfect AI is utilized more within high-stakes contexts, such as decision-making in the medical sector, research has focused on the impacts of potentially inaccurate AI advice on humans' decision-making \cite{kocielnik2019will, lee2019egoistic, robbemond2022understanding}. 
However, there are very few works investigating how \textbf{imperfect XAI} impacts humans' appropriate reliance on AI advice~\cite{papenmeier2019model}.

\begin{figure}[!b]
  \centering
  \includegraphics[width=0.9\linewidth,clip]{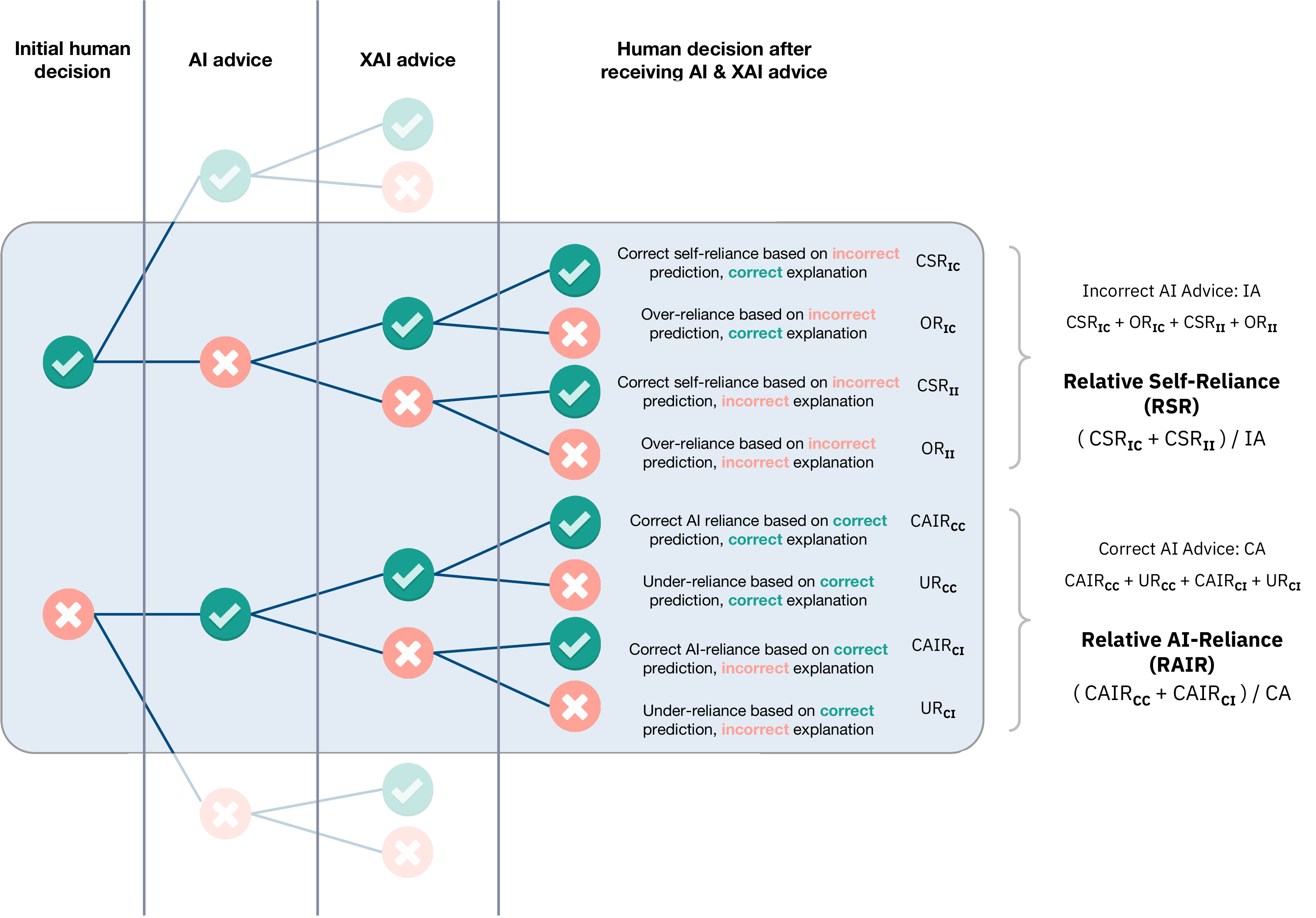}
  \caption{Different paths that human decision-makers could follow based on receiving AI and XAI advice. This figure expands that presented by ~\citet{schemmer2023appropriate} by contributing the XAI advice dimension. The XAI advice is simplified into correct and incorrect explanations. The green checkmarks represent correct advice/decisions, while the red `x` represents incorrect advice/decisions.
  } 
  \label{reliance-model}
\end{figure}

Thus, in this work, we draw from the conceptualization on appropriate reliance previous research has established \cite{bansal2021does, schemmer2022meta, schemmer2023appropriate,schoeffer2023interdependence}. We specifically build on the conceptualization presented by ~\citet{schemmer2023appropriate} in \cref{reliance-model} by adding a new dimension to consider when investigating appropriate reliance in human-AI collaboration: The correctness of XAI advice. We simplified the correctness of XAI advice to be a binary case of correct or incorrect in \cref{reliance-model}.

The introduction of this new dimension unveils previously unexplored avenues within the realm of human-AI collaboration in CSCW, thereby offering a conceptual framework to delve into a more profound comprehension of human decision-making behavior with an AI collaborator. As a result, researchers can calculate more specific metrics regarding human decisions after receiving the AI and XAI advice. For example, the ratio of under-reliance based on a correct prediction and incorrect explanation could be different than when based on a correct prediction and correct explanation. These types of scenarios should not be overlooked when investigating human-AI collaborations in work settings. In \Cref{formulas}, we provide additional details of the newly introduced metrics.

\begin{figure}[!htp]

  \includegraphics[width=0.9\linewidth,clip]{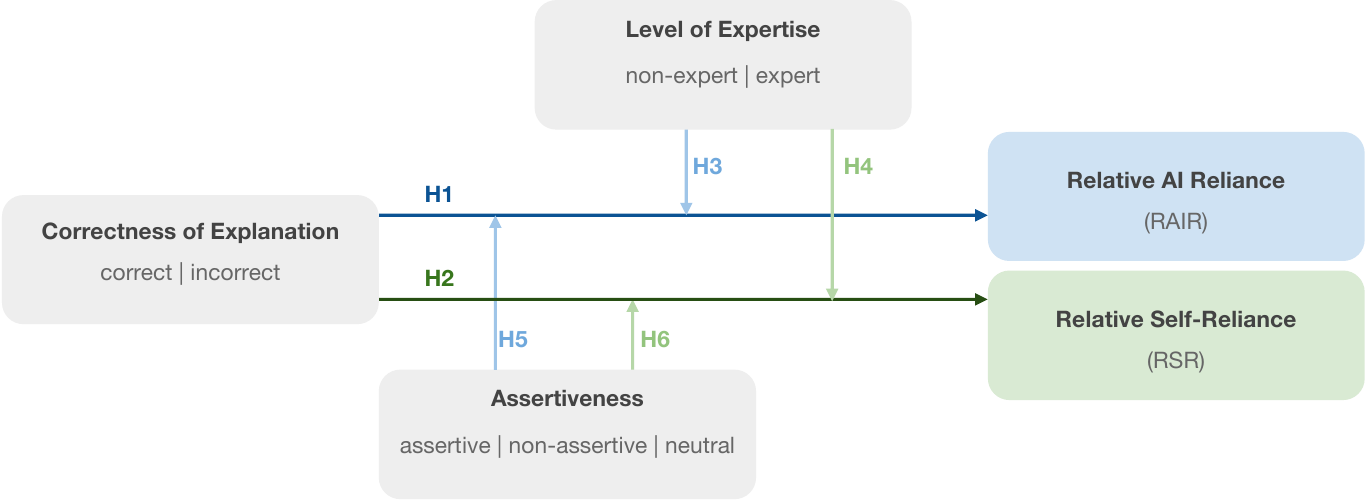}
  \caption{Research model for collaborating with imperfect XAI systems. We analyze the moderation of the level of expertise and assertiveness on the effect of the correctness of explanation on RAIR and RSR.}
  \label{research model}
\end{figure}

With this new dimension for analyzing human-AI collaborations, we investigate the effect that imperfect explanations have on humans' decision-making; we investigate this relation in a sequential decision-making scenario. Based on the constructs of relative AI reliance (RAIR) and relative self-reliance (RSR), we account for the appropriateness of reliance \cite{schemmer2023appropriate}\footnote{Appropriateness of reliance is the quantitative measurement for appropriate reliance. These terms will be used interchangeably throughout the article.}. RAIR comprises the cases in which the human corrects their initially incorrect decision by overriding it with the correct AI advice. On the other hand, RSR comprises all cases in which the human initially makes a correct decision, the AI system gives incorrect advice,  and the human rightly dismisses this advice. Thus, we use appropriateness of reliance as the dependent variable in our research model (see \cref{research model}). In the recent work of \citet{schoeffer2022explanations} the authors investigate how explanations affect distributed fairness in AI-assisted decision-making. Their study shows that task-relevant explanations impact humans' reliance behavior into increasing stereotype-based errors. We apply these findings to our research model and assume that for the cases in which the AI provides correct advice, explanations will affect RAIR. Accordingly, we hypothesize:

\begin{hyp}[resume, wide, leftmargin=0cm, labelindent=0pt, labelwidth=0em]
\item The correctness of explanations impacts humans' relative AI reliance in human-AI decision-making. 
\label{hyp1}
\end{hyp}

\begin{hyp}[resume, wide, leftmargin=0cm, labelindent=0pt, labelwidth=0em]
\item The correctness of explanations impacts humans' relative self-reliance in human-AI decision-making. 
\label{hyp2}
\end{hyp}

One crucial factor in this interrelation between imperfect explanations and humans' appropriate reliance is the level of domain knowledge that humans possess. Previous work shows that humans' level of expertise can influence their decision-making \cite{calisto2023assertiveness, bayer2021role}. Related research in information systems investigates the role of domain knowledge in decision-making. \citet{erjavec2016impact} show in their behavioral experiment in online supply chains that domain knowledge positively impacts humans' confidence in decision-making. Similarly, \citet{dikmen2022effects} analyze humans' reliance on AI when possessing different levels of domain knowledge. In their study, the authors provide an imperfect AI and argue that higher domain knowledge leads to less trust in AI. With this impact of domain knowledge on human-AI decision-making, we intend to examine how the effect of imperfect explanations on appropriate reliance is influenced by humans' level of expertise. Humans with high domain-specific knowledge demonstrate an enhanced ability to discriminate between erroneous explanations and accurate ones with greater rigor \cite{levy2021assessing}. This discernment is facilitated by their extensive expertise, which empowers them to readily identify and discern false information \cite{barfield1986expert}. Thus, in our study, we hypothesize:

\begin{hyp}[resume, wide, leftmargin=0cm, labelindent=0pt, labelwidth=0em]
\item Humans' level of expertise moderates the effect of the correctness of explanations on RAIR.
\label{hyp3}
\end{hyp}

\begin{hyp}[resume, wide, leftmargin=0cm, labelindent=0pt, labelwidth=0em]
\item Humans' level of expertise moderates the effect of the correctness of explanations on RSR.
\label{hyp4}
\end{hyp}

Humans' information processing is not only influenced by what they are provided but also by the way this information is provided. Previous research demonstrates that language style can impact humans' decision-making behavior~\cite{huang2022and,kronrod2022think,calisto2023assertiveness}. \citet{huang2022and} show that the level of assertiveness in reviews affects humans' online review persuasion. Similarly, \citet{kronrod2022think} investigates the effect of the level of assertiveness on the tone of language. They find that there is a relationship between the tone of language and its level of assertiveness. Next to the psychological research field, recent research in HCI examines how assertiveness in explanations affects humans' performance when interacting with AI \cite{calisto2023assertiveness}. They show that the level of assertiveness does impact humans' trust when collaborating with AI. In their work, they reveal that humans are more likely to follow AI advice when the explanation is presented in their own communication style. With those findings and related research on the impact of assertiveness on humans' decision-making, it is reasonable to hypothesize that assertive explanations will impact humans' reliance on AI. Hence, we hypothesize:

\begin{hyp}[resume, wide, leftmargin=0cm, labelindent=0pt, labelwidth=0em]
\item Explanations' level of assertiveness moderates the effect of the correctness of explanations on RAIR. 
\label{hyp5}
\end{hyp}

\begin{hyp}[resume, wide, leftmargin=0cm, labelindent=0pt, labelwidth=0em]
\item Explanations' level of assertiveness moderates the effect of the correctness of explanations on RSR. 
\label{hyp6}
\end{hyp}

Our research model shown in \cref{research model} summarizes the hypotheses that we test in our mixed-methods study. Overall, with this research model, we test for moderating effects of the level of expertise and level of assertiveness on the effect of the correctness of explanations on appropriate reliance.

\section{Methodology}
\label{methodology}
In this section, we describe the task of bird species identification, the experiment design, the recruitment process of participants, and the development of the explanations we use in the study. Finally, we end this section by outlining the data we select to use for the study and the metrics we use to analyze the results.

\subsection{Task Domain: Bird Species Identification}

Computer-supported cooperative work is becoming core to wildlife conservation efforts~\cite{tuia2022perspectives,bondi2022role,green2020innovations}.
With mobile devices becoming increasingly powerful, non-experts and experts alike can use AI-powered applications like the Merlin Bird ID app~\cite{merlin} to identify bird species for monitoring biodiversity and learning about birds. 
The popularity of both birding and AI-based image classification techniques suggests that bird species identification would be a sensible domain to investigate our research techniques.
Furthermore, this is a task for people with a wide range of expertise. 

While identifying bird species from images may not be posed as a high-stakes task in our study, this task is imperative to conserving and managing species and biodiversity~\cite{akccay2020automated}. Furthermore, the task of fine-grained image classification, such as bird species identification, is comparable to higher-stakes tasks, such as identifying diseases from medical imagery~\cite{tschandl2020human}. For example, radiologists collaborating with an imperfect AI and imperfect XAI to diagnose diseases present in chest X-rays would go through a similar visual decision-making process as if they were trying to classify an image of a Bewick Wren in our study interface. ~\citet{kayser2022explaining,hou2021ratchet} propose an imperfect natural language explanation for chest x-rays, similar to the explanation we implement in our study, which helps bridge our findings between bird species identification and higher-stakes tasks.

Previous studies that focused on human-centered XAI and human-AI collaboration also use the domain of bird species identification to understand better human-AI collaboration (\emph{e.g.}, ~\cite{kim2023help,nguyen2022visual,cabrera2023improving}). However, few previous works focus on human-AI collaboration for decision-making in the wildlife conservation domain overall. Yet, the field of AI for wildlife conservation is rapidly growing~\cite{tuia2022perspectives} and could benefit from research related to CSCW and HCI.

\subsection{Study Design}
\label{study-design-section}

\begin{figure}[!b]
  \centering
  \includegraphics[trim={0cm 8cm 12cm 0cm},width=\linewidth,clip]{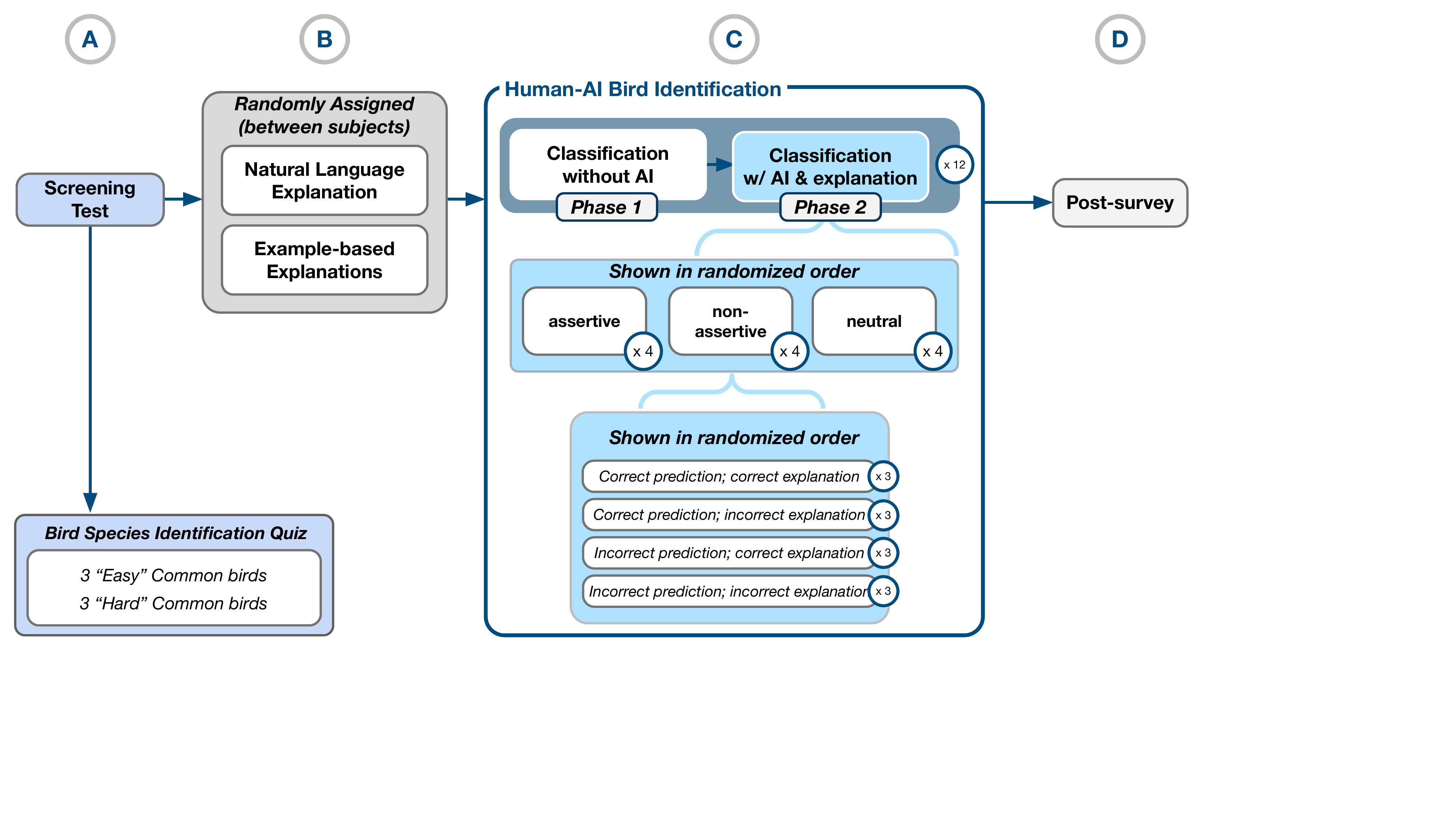}
  \caption{We conduct a rigorous mixed-methods study leveraging a mixed design. Before participants start the task, they are shown a screening test (A). For the human-AI bird identification task, participants are assigned an explanation modality (B). During the task, participants are shown explanations with different levels of assertiveness and different scenarios of correctness (C). Lastly, participants complete a post-survey (D).}
  \label{experimentdesign}
\end{figure}

To answer our research questions, we design a mixed between- and within-subjects study to examine various effects of explanations on appropriate reliance. Our study was carefully reviewed by two experienced birders\kmedit{: one experienced birder is a migration counter for a bird sanctuary, and the other experienced birder holds a graduate degree in environmental science, conducted research at a nature center, and worked at a nature conservancy}. The procedure of the study follows the design outlined in \Cref{experimentdesign} and is divided into four different parts ($A-D$), which we explain in further detail below. 

The study begins with part A in \cref{experimentdesign}: In a bird identification test, we assess participants' expertise in classifying six different species of birds\footnote{Specific birds used for the bird identification test are reported in \cref{bird-test-details}, p. \pageref{bird-test-details}.}. We distinguish the six bird images based on their level of difficulty. This level of difficulty is derived from discussions with an experienced migration counter from a bird sanctuary. \kmedit{While previous work has collected participants' self-perception of expertise~\cite{kim2023help}, this method is subjective, and participants may self-perceive their skills differently. ~\citet{kazemitabaar2023novices} measure participants ``experience-level'' through log data instead of subjective measures. Similarly, we try to avoid defining expertise subjectively.}
For the purpose of our analyses, we identify two different levels of expertise: \textit{non-experts} and \textit{experts}.

In the next section of the study (part B in \cref{experimentdesign}), participants are randomly assigned to one of two explanation types. Similar to previous studies \cite{riefle2022influence, robbemond2022understanding, szymanski2021visual, chandrasekaran2017takes}, the treatments differ in the explanation modality participants receive: \textit{Natural language} explanations or \textit{visual, example-based} explanations. We use natural language explanations because the AI model that we used for the study was specifically designed to generate natural language explanations based on fine-grained image classifications~\cite{hendricks2021generating}. We choose to also look at example-based explanations as recent studies focus on this modality in human-AI collaboration~\cite{chen2023understanding,kim2023help,cai2019effects,humer2022comparing}. \psedit{However, recent studies show that example-based explanations have potential benefits. \citet{chen2023understanding} show that example-based explanations improve humans' performance, so much so that it leads to complimentary team performance. With promising results from previous research and numerous clinical decision-support tools proposing to incorporate example-based explanations (\emph{e.g.}, ~\cite{sadeghi2018users,barata2021improving}), we find it necessary to investigate the effect of example-based explanations on humans' appropriate reliance in the context of imperfect XAI.}

\begin{figure}[!h]
  \centering
  \includegraphics[trim={-3cm 10cm 3cm 0cm},width=\linewidth,clip]{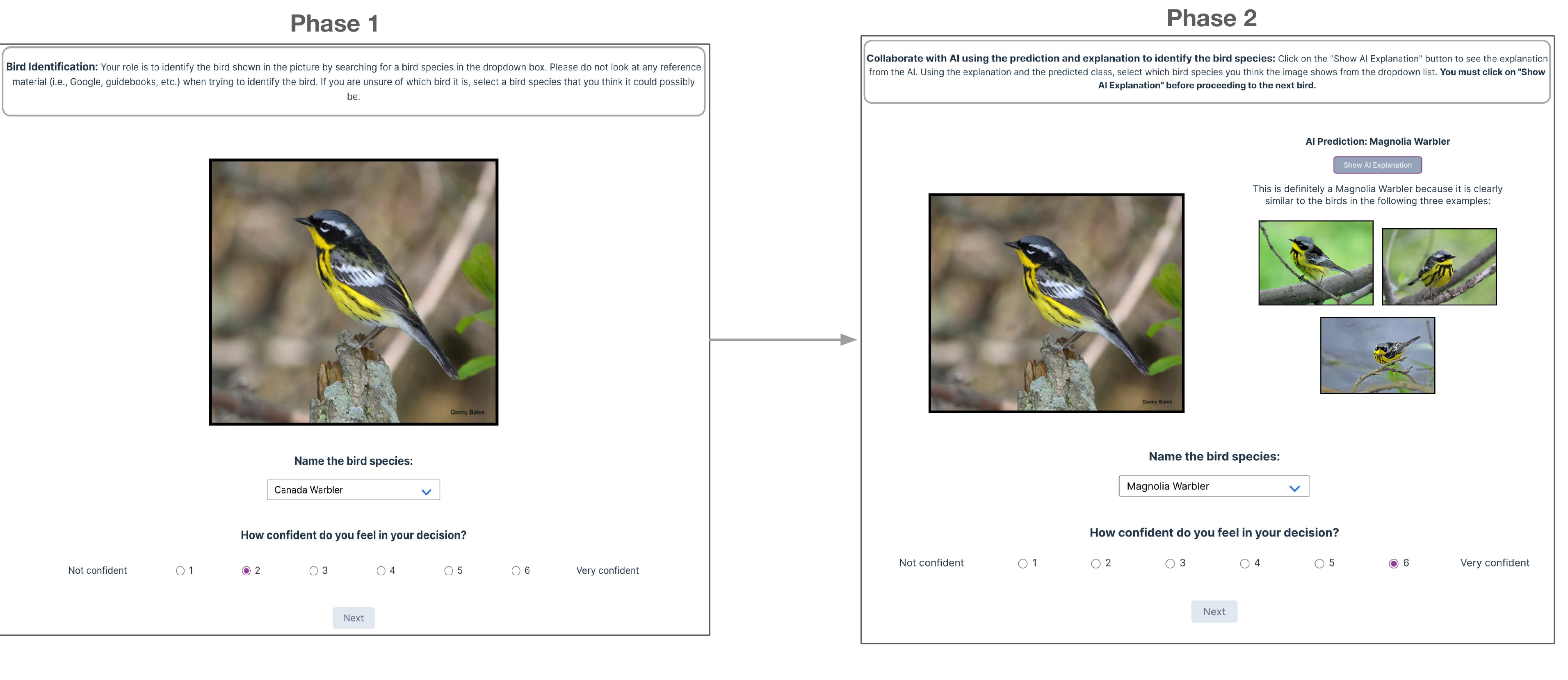}
  \caption{Example of the two phases for a single bird image that a participant is shown in the study. This specifically shows a Magnolia Warbler (correct prediction, correct explanation), and this participant is assigned to the example-based explanations. For this bird, the participant is shown an assertive explanation. }
  \label{experimentphases}
\end{figure}

The human-AI bird identification task (part C of \cref{experimentdesign}) consists of two phases. For each treatment condition, the participants are asked to initially identify the bird species from an image (phase 1 in \cref{experimentphases}). After submitting an initial identification, they are shown the AI's prediction along with the explanation, and again, they have to submit an identification for the bird species in the image (phase 2 in \cref{experimentphases}). The structure of phases one and two are corroborated with previous work~\cite{green2019principles}. Initially, participants must click on a button that shows ``Show AI Explanation''. Without revealing the explanation, the participant cannot proceed to the next question. This is one way for us to ensure that the participant acknowledges the presence of an explanation. Overall, participants do this process for twelve different random bird images.

As the AI that we are utilizing for identifying the bird species is not perfect~\cite{hendricks2021generating}, the predictions and explanations provided can be incorrect. In order to understand how this affects participants' appropriate reliance, we ensure that each participant is shown three samples of the following four categories in random order: 
\begin{itemize}
    \item \textbf{CC}: correct prediction and correct explanation
    \item \textbf{CI}: correct prediction and incorrect explanation
    \item \textbf{IC}: incorrect prediction and correct explanation
    \item \textbf{II}: incorrect prediction and incorrect explanation
\end{itemize}
Overall, participants are shown twelve different bird species. For each of the four categories we identified, we show three samples where each explanation is framed with a different level of assertiveness: \textit{assertive}, \textit{non-assertive}, or \textit{neutral}\footnote{\Cref{explanationformat} provides examples of \textit{assertive}, \textit{non-assertive}, and \textit{neutral} explanations.}. As a result, each participant is shown one \textit{assertive}, one \textit{non-assertive}, and one \textit{neutral} explanation for each category. We ensure that the order is randomized for each participant. Moreover, we also vary the samples shown, meaning that not every participant sees the same bird images. This is to ensure that our results are not dependent on the difficulty of the bird species.

After finishing the task, participants must fill out an additional questionnaire (part D of \cref{experimentdesign}). Here, we qualitatively assess participants' ability to properly rely on the AI based on the explanations that they were shown. Thus we ask them: ``\textit{Under what circumstances would you prefer assertive (e.g., “definitely”, “clearly”) versus non-assertive (e.g., “might be”, “appears to be”) versus neutral explanations and why?}''. Aside from this open-text question, we also ask participants about their occupations and the regions in North America that they are most familiar with in terms of bird species.


\subsection{Data Selection}

We create a dataset of bird images and explanations to show participants by manually curating bird images from the well-established CUB-200-2011~\cite{WahCUB_200_2011} dataset and explanations from the Generating Visual Explanations model~\cite{hendricks2021generating}. The original dataset consists of 11,788 images of 200 different bird species and is split into 5,994 training and 5,794 test images. Each bird species is represented with around 60 images of the respective bird class. When curating birds, we first filtered for bird families with several species in the CUB dataset. We specifically filtered out every bird class that is not a part of the Warblers, Wrens, Swallows, Sparrows, or Finches/Grosbeaks families. After applying this filter, we had $1,864$ out of $5,794$ images from the test set of CUB-200-2011. Of those $1,864$ images, $1,609$ were predicted correctly by the AI, and $255$ images were predicted incorrectly by the model. 

After filtering the bird species, multiple researchers on our team separately classified the natural language explanations and the visual, example-based explanations for a subset of the $1,864$ birds as incorrect or correct. Cases of doubt were discussed by a subset of the research team and excluded from consideration if an agreement was not met. In total, we identified ten examples for each category\footnote{The four categories are identified in Section \ref{study-design-section}.} and explanation type. In some cases, the example-based and natural language explanations for a single bird are used. As a result, the dataset represents 66 different images and 43 different bird species from the CUB-200-2011 dataset. 

\begin{figure}[!hbt]
  \centering
  \includegraphics[trim={3cm 0cm 8cm 0cm},width=\linewidth,clip]{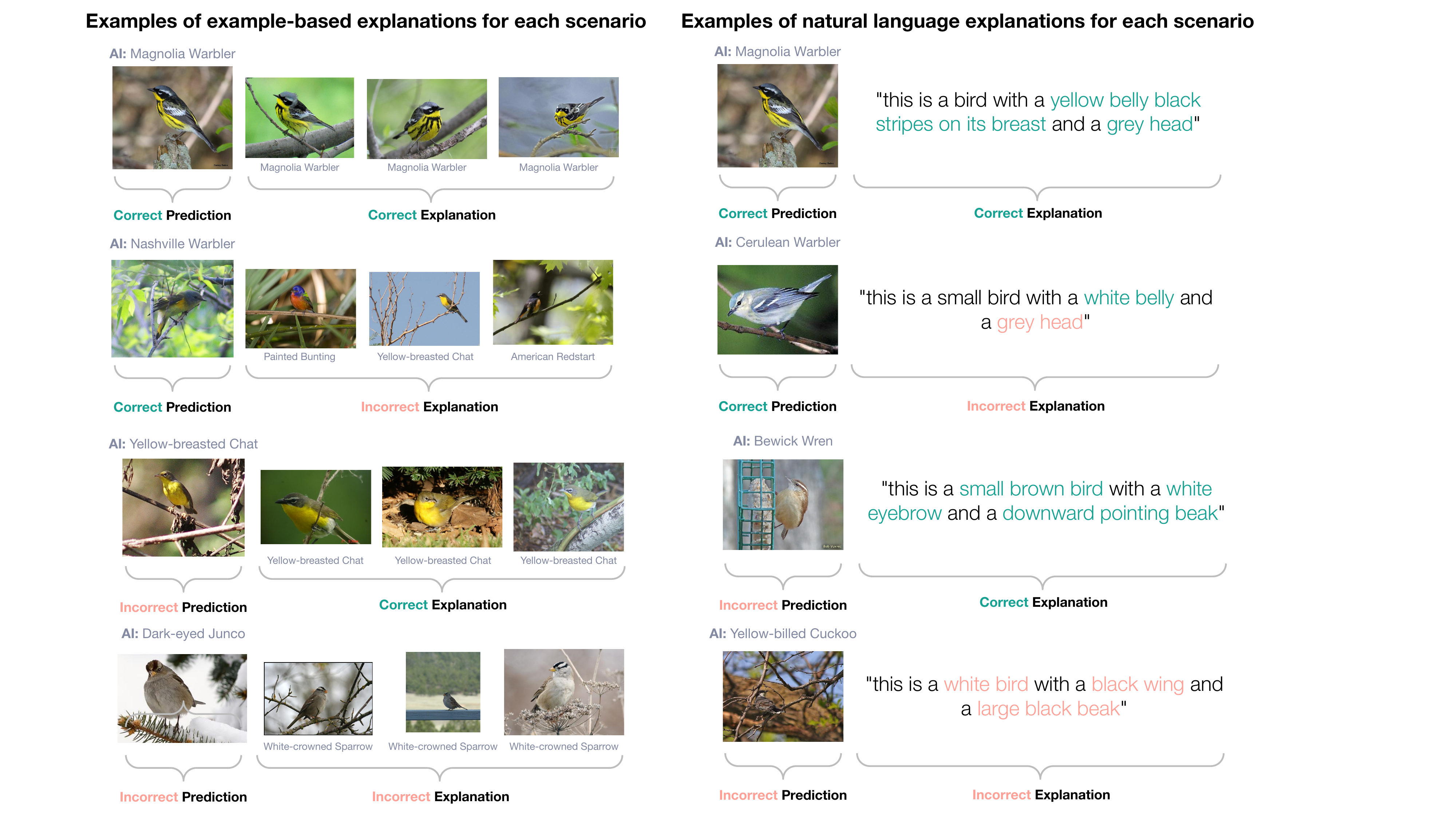}
  \caption{Representative examples of the example-based and natural language explanations for each scenario: CC, CI, IC, and II. The class of the example-based images in the explanation is not shown to participants during the study. The red and green coloring on the natural language explanations was not shown during the study. This is only provided in the figure to guide the reader. The natural language explanation for the Cerulean Warbler is incorrect because this bird species does not have a grey head. The natural language explanation for the Yellow-billed Cuckoo is incorrect because this bird species is brown with a white belly, has a gold and black beak, and does not have a black wing. }
  \label{explanationexamples}
\end{figure}

We define a correct natural language explanation to be when the explanation aligns with the description of the predicted bird class. We define an incorrect natural language explanation to misalign with all or part of the description of the predicted bird class. \psedit{Thus, an incorrect natural language explanation contains a factual error. This type of incorrectness is present in different natural language techniques and is a focus of current research \citep{liu2022improving, xie2023faithful}.} We use descriptions from the Cornell Lab of Ornithology All About Birds Guide~\cite{birdsguide} to corroborate our classification for each explanation. Examples of incorrect and correct natural language explanations are provided in \cref{explanationexamples}. 

For the example-based explanations, we define a correct explanation as the three most similar images belonging to the predicted class (as shown in phase 2 of \cref{experimentphases}). We define an incorrect explanation to be most similar to at least one image that is not of the predicted class. \psedit{This means that incorrect example-based explanations incorporate logical errors as the examples shown are dissimilar from the predicted class. Moreover, such incorrect explanations can have an inconsistency as the examples shown might differ in the classes shown.} However, we only choose to show participants incorrect explanations that have at least two images that are not of the predicted class. For example, in \cref{explanationexamples}, the AI correctly predicts a Nashville Warbler; however, the three most similar examples are a Painted Bunting, a Yellow-Breasted Chat, and an American Redstart.  In some cases where the advice is correct and the explanation is incorrect\footnote{The explanation does not align with the predicted class.}, the explanation may align with the ground truth class. For example, for a Tennessee Warbler, the AI predicts an Orange-crowned Warbler (incorrect advice since the wrong bird species is predicted), but the three most similar examples are all of Tennessee Warblers (incorrect explanation since the examples' bird species do not align with prediction). It's possible that a model could be relying on spurious patterns to make classifications~\cite{plumb2021finding}. Since we are dealing with an imperfect AI, we do not choose to exclude such cases from our dataset.

\subsection{Explanation Modalities}

\paragraph{Natural Language Explanations}
The natural language explanations were generated by the model proposed by Hendricks et al.~\cite{hendricks2021generating}. We followed the PyTorch implementation~\cite{salanizpytorch-gve-lrcn} of Hendricks et al.'s model to obtain the natural language explanations since the original model from Hendricks et al. was unavailable. 
After running the test images through the model, a natural language explanation is generated for each classification. For example, the natural language explanation for the Magnolia Warbler in \cref{explanationexamples} is: ``\texttt{this is a bird with a yellow belly black stripes on its breast and a grey head}''.  

\paragraph{Example-Based Explanations}
Previous work creates example-based explanations, specifically normative explanations, by calculating the Euclidean distance between the given image and the images in the dataset~\cite{cai2019effects}. Another study generates the example-based explanation by calculating the $L_{2}$ distance of the embedded features~\cite{nguyen2021effectiveness}. 
We generated the example-based explanations by following methods used in previous works~\cite{barata2021improving,tschandl2020human}. As done by \citet{tschandl2020human} and \citet{barata2021improving}, we calculate the cosine similarity between the extracted feature vector of the given image and the rest of the extracted feature vectors of the images in the training set. Unlike Barata and Santiago~\cite{barata2021improving}, we choose not to take the example's ground truth class into consideration. The extracted features from the images were provided by Hendricks et al.~\cite{hendricks2021generating}. Because the model is not perfect, the example-based explanations are also not perfect. For example, even though the model correctly predicted an image of a Nashville Warbler in \cref{explanationexamples}, the three most similar images are of three different birds. For this study, we consider an example-based explanation to be incorrect if two of the three examples are of a different class than the predicted class.
Inspired by Ford et al., we choose to show participants the three most similar examples~\cite{ford2022explaining}.

\begin{figure}[!b]
  \centering
  \includegraphics[trim={0cm 15cm 0cm 0cm},width=\linewidth,clip]{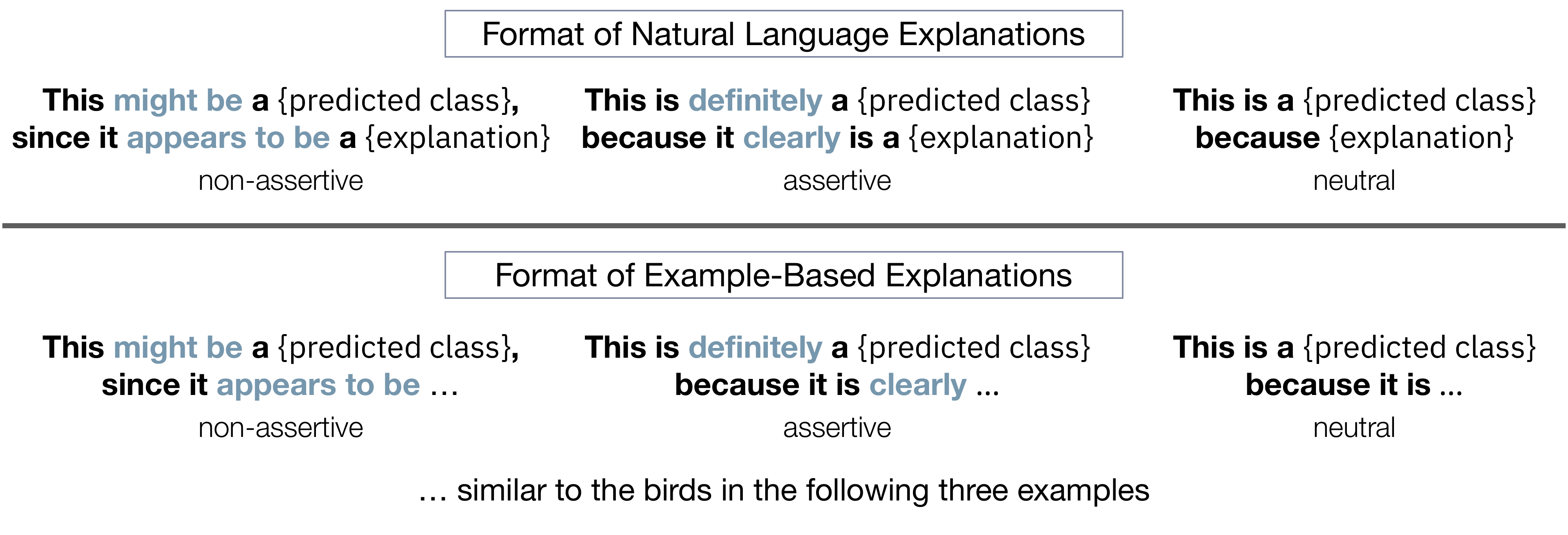}
  \caption{The three different language tones that an explanation could have in terms of assertiveness for the natural language and example-based explanation modality. Non-assertive explanations included the words ``might be'' and ``appears to be''. Assertive explanations included the words ``definitely'' and ``clearly''. Neutral explanations did not include any additional adjectives.}
  \label{explanationformat}
\end{figure}

\paragraph{Assertiveness of Explanations} Following previous studies~\cite{winter2015don,calisto2023assertiveness,pacheco2019alignment}, we define assertive explanations to include words and adjectives such as ``definitely'' and ``clearly''. We define the non-assertive explanations to include words and adjectives such as ``might be'' and ``appears to be''. For the neutral condition, we omit the adjectives to maintain the same structure of the information being presented. For the natural language explanations, we append the assertiveness to the beginning of the explanation generated by the model to read like a sentence. For the non-assertive and assertive conditions, we removed the text ``this is a'' from the generated explanation in order to incorporate it into the sentence structure we designed. The three versions of assertiveness for both explanation modalities are shown in \cref{explanationformat}. To our knowledge, there is no literature to rationalize how to appropriately present assertiveness visually for example-based explanations, so we opt to use natural language in combination with the example-based explanations. 

\subsection{Recruitment}

We recruit the participants through several communication channels that are related to the environment and conservation, such as the AI for Conservation Slack, Birding International Discord, Climate Change AI community forum, WildLabs.net community forum, and Audubon Society mailing lists. Additionally, we use Prolific as previous research has indicated that this platform is a reliable source of research data \cite{peer2017beyond, palan2018prolific}. We apply a custom filter on Prolific to target individuals who currently work in a field related to nature, science, the environment, or animals. Participants receive compensation that is above minimum wage.
Overall, we try to limit recruitment to only address people with prior knowledge of birding to minimize the prevalence of novices' randomly guessing bird species identification. 
After excluding participants who provide incomplete and fake responses (i.e., lorem ipsum response to our survey question), we have $136$ people complete our study. In order to determine if a participant is familiar with birding, participants take a bird identification test (phase A in \cref{experimentdesign}). The participant's score on the bird identification test is used to determine whether they are considered a non-expert or an expert. Details related to clustering participants based on their test scores are provided in \cref{results_section}.

\subsection{Quantitative and Qualitative Metrics}

\paragraph{Quantitative Metrics} We quantitatively calculate appropriate reliance across the four dimensions defined by \citet{schemmer2023appropriate}: correct AI reliance, correct self-reliance, under-reliance, and over-reliance. Accordingly, correct AI reliance measures the number of correct decisions when the human's initial decision is incorrect, and the human is rightly taking over the correct AI advice. Correct self-reliance is when the human initially makes the correct decision and does not overwrite their decision with the incorrect AI advice. On the other hand, under-reliance reflects the case in which the human initially makes an incorrect decision and does not adhere to the correct AI advice. On the other hand, over-reliance represents the scenario in which the human makes an initial correct decision but overrides her own decision with incorrect AI advice. Following the appropriate reliance metrics defined by ~\citet{schemmer2023appropriate}, we calculate RSR and RAIR to account for the appropriateness of reliance. 

With the new dimension for XAI advice, we can separately measure RAIR and RSR for correct and incorrect explanations and derive its impact on appropriate reliance. In order to measure this impact, we look at the Deception of Reliance (DoR) caused by imperfect XAI. For RAIR, we can apply the following: 
\begin{equation}
    DoR_{RAIR} = RAIR_{C} - RAIR_{I}.
\end{equation}
In this equation, the subscript $I$ represents incorrect explanations, whereas the subscript $C$ represents correct explanations. We can compute the same for RSR: 
\begin{equation}
    DoR_{RSR} = RSR_{C} - RSR_{I}.
\end{equation}
In order to measure the overall deception impact of explanations on humans' decision-making behavior, we compute the deception on appropriate reliance by calculating the Gaussian distance in the RAIR-RSR space between incorrect and correct explanations:
\begin{equation}\label{DAIR_aor_eq}
    DoR(RAIR, RSR) = \sqrt{{(RAIR_{C} - RAIR_{I})}^2 + {(RSR_{C} - RSR_{I})}^2 }.
\end{equation}
According to Schemmer et al.'s conceptualization of Appropriateness of Reliance \cite{schemmer2023appropriate}, this results in the following:
\begin{equation} 
    DoR_{AoR} = AoR_{C}(RAIR, RSR) - AoR_{I}(RAIR, RSR).
\end{equation}

This difference represents the deception between the correct and incorrect explanations. If the deception is a positive value, then incorrect explanations are more deceptive; if the difference is a negative value, then correct explanations are more deceptive. 

Lastly, as defined by previous work (\emph{e.g.}, ~\cite{hemmer2021human,bansal2021does,green2019principles}), we can calculate the human-AI team performance to determine if CTP exists. Following the constructs defined in those previous works, we determine if CTP exists by calculating the participants' performance in identifying the bird species \textbf{before} and \textbf{after} they see the AI advice and compare this to the performance of the model on the twelve birds images shown to the participant. We utilize accuracy as a performance metric. Since every participant is shown six birds that the AI correctly classifies and six that the AI incorrectly classifies, the model performance is $50\%$.  

\paragraph{Qualitative Metrics} We also conduct an inductive content analysis of the open-ended responses to better understand the participants' preferences regarding assertiveness. As a reminder, in the end-survey, we ask participants specifically ``\textit{Under what circumstances would you prefer assertive (e.g.,
``definitely'', ``clearly'') versus non-assertive (e.g., ``might be'', ``appears to be'') versus neutral explanations
and why?}''. To analyze these responses, we follow the established procedure of \citet{gioia2013seeking} and screen the answers in three coding workshops. In the first workshop, the first and second authors of this article initially screen the answers and applied open coding \cite{holton2007coding} to extract and aggregate core constructs of participants' answers. In this procedure, both authors discuss their findings and align their understanding of relevant constructs. Through a second coding workshop, we apply axial coding to derive subcategories of these constructs and align those with the data. In a final workshop, we distill those emerging themes and derive aggregated dimensions \cite{wolfswinkel2013using}. 
\section{Results}
\label{results_section}

To answer our research questions and confirm or reject our hypotheses, we conduct rigorous statistical and qualitative analyses. We measure appropriate reliance based on metrics defined in previous work: Relative AI reliance (RAIR) and relative self-reliance (RSR)~\cite{schemmer2023appropriate}. By doing so, we answer \textbf{RQ} \ref{rq1} and \textbf{RQ} \ref{rq2} in \Cref{section_ar}. We also calculate the participant's task accuracy before and after receiving AI and XAI advice to answer \textbf{RQ} \ref{rq3} (see \Cref{hai_performance}). This way, we can determine whether complementary team performance exists~\cite{hemmer2021human}. Additionally, based on the new metric, Deception of Reliance (DoR), we measure to what extent explanations deceive humans, answering \textbf{RQ} \ref{rq4} in \Cref{deception_section}. Lastly, we qualitatively analyze open-ended responses through rigorous inductive content analysis in \Cref{qual-res}. For all of our research questions, we look at two different types of explanations: natural language explanations that are focused on specific features present in the image and visual, example-based explanations showing the top three most similar example images from the training set. While our analyses look at both modalities, we do not intend to compare them directly. Therefore, we do not conclude one modality is better or worse than the other.

\subsection{Participant Statistics}
On average, the study takes $24$ minutes to complete.
In order to distinguish experts from non-experts, we perform K-means clustering ($k=2$) based on a principal component analysis with two components for four features from the bird species identification test (part A of \cref{experimentdesign}). These four features represent participants' scores in correctly identifying the family and species of the easy and the difficult bird images. By clustering the $136$ participants into the expert and non-expert group, we end up with $83$ experts and $53$ non-experts. With this clustering, the average bird identification test score (summing up all four scores in the identification test) for non-experts is $38.99\% (STD = 11.42\%)$ while the average test score for experts is $83.84\% (STD = 12.30\%)$\footnote{Participants performance on the bird identification test is shown in \cref{test-scores}, Appendix Section \ref{bird-test-details}.}. Of the $83$ experts, $42$ see example-based explanations, and $41$ see natural language explanations. Of the $53$ non-experts, $25$ see example-based explanations, and $28$ see natural language explanations. In terms of the fields that the $136$ participants represent, $45$ participants have an occupation primarily related to biology, conservation, and/or the environment. $26$ have an occupation primarily related to engineering and/or technology; $30$ are either researchers, students, or affiliated with education in some other way; $24$ have occupations in miscellaneous industries; and $11$ are retired.

\subsection{Moderating Effects in Imperfect XAI Research Model}
\label{section_ar}

In order to test whether humans' level of expertise and the explanations' assertiveness moderate the relation of the correctness of explanations on humans' appropriate reliance, we conduct several moderation analyses utilizing the process macro model of \citet{hayes2017introduction}. 
An overview of the regression analyses is presented in \Cref{mod_anal}.

\begin{table}[htbp!]

\caption{Moderation analyses of the correctness of natural language and example-based explanations on RAIR and RSR with the level of expertise and assertiveness as moderators. The coding of assertiveness used for the moderation analyses is provided. }

\begin{tabular}{P{2cm} P{1cm} P{1cm}}
\multicolumn{3}{c}{Coding of assertiveness} \\

\hline

assertiveness &  Z1 & Z2 \\ \hline \hline

\textit{neutral} & 0 & 0 \\  \hline
\textit{non-assertive} & 1 & 0 \\  \hline
\textit{assertive} & 0 & 1 \\  \hline
\\\\ \end{tabular}

\begin{threeparttable}

\begin{tabular}{m{1.5cm}R{0.9cm} R{0.9cm} R{0.002cm} R{0.9cm} R{0.9cm} R{0.002cm} R{0.9cm} R{0.9cm} R{0.002cm} R{0.9cm} R{0.9cm}} \hline
& \multicolumn{5}{c}{RAIR} && \multicolumn{5}{c}{RSR}\\
\cmidrule{2-6} \cmidrule{8-12}
& \multicolumn{2}{c}{Natural Language} &&  \multicolumn{2}{c}{Example-Based} & & \multicolumn{2}{c}{Natural Language} & & \multicolumn{2}{c}{Example-Based}  \\
\cmidrule{2-3} \cmidrule{5-6} \cmidrule{8-9} \cmidrule{11-12}
& coeff & p&  & coeff & p & & coeff & p&  & coeff & p \\
\hline \hline
const   & 1.26   & .00 & & .43 & .17 & & -17.16   & .98  & & -3.60  & .00    \\\hline
corr   & .57   & .29 & & 1.02 & .04 & & 13.25   & .98  & & -.4.36  & .74    \\\hline
exp   & 2.12  & .00 &&  -1.25 & .00 &  & 15.89   & .98 &  & 3.25  & .00    \\\hline
Z1   & -.46 & .24 &&  .26 & .47 &&  .14   & .26  & & -.09  & .83    \\\hline
Z2   & .00  & 1.00 &&  .00 & 1.00 & & -.31   & .58  & & .09  & .83    \\\hline 
exp x corr   & -1.00  & \best{.05} & & -1.04 & \best{.03} & & -13.33  & .98  & & -.73  & .57    \\\hline
Z1 x corr   & .46  & .44 &&  -.03 & .95 & & -.28   & .71  & & -.45  & .55    \\\hline
Z2 x corr   & .19  & .74 &&  -.16 & .77 & & .90   & .23  & & -.43  & .55   \\\hline
\end{tabular}
    \begin{tablenotes}
        \item[1] \textit{corr} --- \textit{correctness}; \textit{exp} --- \textit{level of expertise}
    \end{tablenotes}
    \end{threeparttable}

\label{mod_anal}

\end{table}

\subsubsection{Participants' level of expertise moderates the effect of the correctness of explanations on RAIR for natural-language explanations}
\label{mod_analysis_nle_rair}
As theoretically developed in Section \ref{theoretical_section}, we model the correctness of explanations as an independent variable. Accordingly, we model RAIR as the dependent variable. To account for the moderation effect of the level of expertise and assertiveness, we examine each variable as a moderator and report the interaction effects with the correctness of explanations. The results of this moderation analysis are shown in \Cref{mod_anal} (a detailed view is shown in \Cref{mod_anal_nle_rair} on p. \pageref{mod_anal_nle_rair} in the \Cref{mod_appendix}).

The moderation analysis shows that the interaction of the level of expertise with the correctness of explanations is significant (coeff $= -1.00$, p-value $= .05$). We observe a negative coefficient. Accordingly, the moderation effect on the relation of correctness on RAIR is higher for non-experts than for experts. \kmedit{In other words, non-experts change their initially incorrect decision to align with the correct AI advice more often than experts do when the natural language explanation is correct.} However, there is no significant effect in the interaction of assertiveness and the correctness of explanations. Thus, we conduct a regression analysis with the moderators as independent variables to evaluate for a direct effect of assertiveness as recommended by \citet{hayes2017introduction} and \citet{warner2012applied}. The results of the regression analysis show that there is no direct effect between assertiveness and RAIR (coeff $= .04$, p-value $= .77$). Thus, we confirm hypotheses \ref{hyp1} and \ref{hyp3} and reject hypothesis \ref{hyp5} for natural language explanations.

\subsubsection{Participants' level of expertise moderates the effect of the correctness of explanations on RAIR for example-based explanations}

Next, we present the moderation analysis of example-based explanations on RAIR. We set up the analysis for example-based explanations the same as the analyses for natural language explanations (see \Cref{mod_analysis_nle_rair}). As seen in \Cref{mod_anal}, there is a significant moderation effect of the level of expertise (coeff $= -1.04$, p-value $= .03$). The negative coefficient signals that this moderation is higher for non-experts than for experts. The correctness of the example-based explanations has a positive coefficient (coeff $= 1.02$, p-value $= .04$), and thus, correct explanations have a positive impact on RAIR. \psedit{Thus, if participants are provided with a correct explanation, they more often correctly follow the AI advice.} \kmedit{Overall, correct explanations result in humans changing their initially incorrect decisions to align with the correct AI advice more often, and this is especially prevalent among non-experts.}

Furthermore, the analysis reveals that assertiveness does not moderate the effect between correctness and RAIR. According to \citet{hayes2017introduction} and \citet{warner2012applied}, we drop the interaction term and conduct a regression analysis with assertiveness set as the independent variable. The result shows that there is no significant direct effect of assertiveness on RAIR (coeff $= -.04$, p-value $= .79$).
Hence, we confirm hypotheses \ref{hyp1} and \ref{hyp3} and reject hypothesis \ref{hyp5} for example-based explanations.

\subsubsection{Participant's level of expertise has a direct effect on RSR for natural language explanations}
\label{mod_analysis_nle_rsr}
In addition to analyzing whether the level of expertise and assertiveness moderate the effect of explanations' correctness on RAIR, we conduct the same analyses for the effect of explanations' correctness on RSR. For RSR, we look at all cases in which the AI prediction is giving incorrect advice (i.e., the prediction is wrong) and the initial human decision is correct \cite{schemmer2023appropriate}. 

The moderation analysis in \Cref{mod_anal} for the natural language explanation shows that there is no significant effect of correctness on RSR moderated by level of expertise (coeff $= -13.33$, p-value $= .98$) and assertiveness (Z1 x corr.: coeff $= -.28$, p-value $= .71$; Z2 x corr.: coeff $= .90$, p-value $= .23$). 

Thus, we perform a regression analysis with the level of expertise and assertiveness as independent variables and drop the interaction terms. We observe that there is no significant effect of assertiveness on RSR (coeff $= .10$, p-value $= .58$). However, the level of expertise (coeff $= 3.18$, p-value $= .00$) has a significant effect on RSR. 
\kmedit{With a positive coefficient, this means that experts dismiss incorrect AI advice more than non-experts when shown natural language explanations.}
\psdelete{The trend} \psedit{This can also be seen} in \cref{rsr-rair}\psedit{, which} tells us that experts have a higher RSR than non-experts. 
Therefore, we reject hypothesis \ref{hyp2}, and additionally, hypothesis \ref{hyp4} as the level of expertise does not have a moderating role but has a direct effect on RSR for natural language explanations. On top of that, we reject hypothesis \ref{hyp6}.

\subsubsection{The correctness of explanations and participants' level of expertise have a direct effect on RSR for example-based explanations}

Lastly, we report the results of the moderation analysis for example-based explanations on RSR. The analysis is set up the same as it is in \Cref{mod_analysis_nle_rsr} but for the example-based explanations. 

We can see in \Cref{mod_anal} that the level of expertise does not significantly moderate the effect of correctness on RSR (coeff = -.73, p-value = .57).
Additionally, there is no significant moderation of assertiveness (Z1 x corr.: coeff $= -.45$, p-value $= .55$; Z2 x corr.: coeff $= -.43$, p-value $= .55$). Thus, we conduct a regression analysis without the interaction terms. The results of this analysis show no direct effect of assertiveness on RSR (coeff $= -.03$, p-value $= .86$). However, there is a significant direct effect of explanations' correctness on RSR (coeff $= -1.40$, p-value $= .00$) and a direct effect of level of expertise on RSR (coeff $= 3.05$, p-value $= 0.00$). \psedit{This means that experts more often correctly override the wrong AI advice and stick to their correct initial decision compared to non-experts for example-based explanations. Additionally, when the explanations are correct, participants more often correctly override wrong AI advice and stick to their correct initial decision.} Hence, experts have a higher RSR than non-experts, which can also be seen in \Cref{rsr-rair}. Moreover, as incorrect explanations have a higher impact on RSR, experts are able to identify false AI advice for incorrect explanations better. This also shows that experts are able to identify incorrect AI advice to a greater extent than non-experts; experts, in this case, rely more heavily on their own judgment.

Thus, we confirm hypothesis \ref{hyp2} and reject hypothesis \ref{hyp4} as the level of expertise does not take in a moderating role but has a direct effect on RSR for example-based explanations. On top of that, we reject hypothesis \ref{hyp6}.
\\
\\
Overall, the moderation analyses reveal that the level of expertise moderates the effect of the correctness of explanations on RAIR for both explanation modalities. Additionally, the analyses show that the level of expertise has a direct effect on RSR for both explanation modalities, and the correctness of explanations has a direct effect on RSR for example-based explanations.

\subsection{Human-AI Team Performance}
\label{hai_performance}
\begin{figure}[!h]
  \centering
  \includegraphics[width=0.9\linewidth,clip]{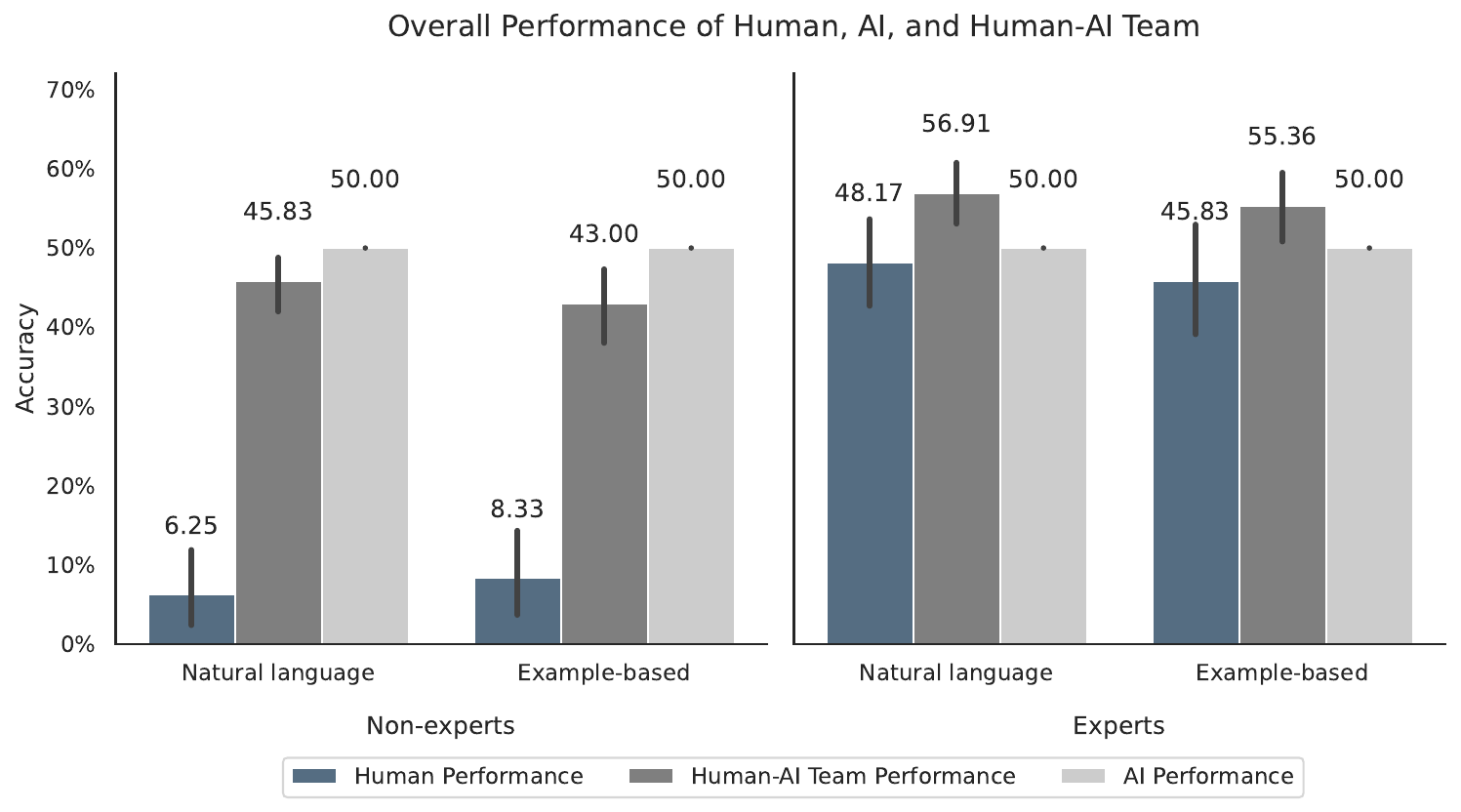}
  \caption{The average overall performance of the human, AI, and human-AI teams for identifying 12 birds. The bar chart on the left shows the performance of the non-experts, while the bar chart on the right shows the performance of the experts.  }
  \label{reliance-counts}
\end{figure}

Hemmer et al. argue that interpretability is a key component of human-AI complementarity~\cite{hemmer2021human}. Several previous user studies have failed to show that incorporating XAI into AI systems can lead to CTP~\cite{fok2023search}. However, with a new dimension of XAI advice in \cref{reliance-model}, we can contribute to the current literature by investigating how the correctness of explanations affects CTP. By calculating the participants' performance before and after seeing the AI and XAI advice, we can determine whether CTP exists in the presence of imperfect XAI. 
As the analyses in \Cref{section_ar} reveal, the level of expertise impacts appropriate reliance in terms of RSR and RAIR. Thus, in comparing the human-AI team performance, we distinguish by participants' level of expertise. We use accuracy as the performance metric. \Cref{reliance-counts} presents the performance of AI and humans for each treatment.

The AI's performance is always $50\%$ because the study was designed to show participants six birds that the model correctly classified and six that the model incorrectly classified. In \Cref{reliance-counts}, we see that when experts are paired with the AI, their performance improves by $8.74\%$ for the natural language modality and $9.53\%$ for the example-based modality. When experts are paired with AI, they perform $6.91\%$ better than the AI alone for natural language explanations and $5.36\%$ for example-based explanations.

While experts reach CTP, we do not see this for non-experts. However, we do see that the non-experts greatly improve their performance and nearly match the AI's performance when paired with the AI. Specifically, non-expert participants who see the natural language explanations improve their performance by $39.58\%$ (task accuracy of $45.83\%$), while non-expert participants who see the example-based explanations improve their performance by $34.67\%$ (task accuracy of $43.00\%$) when paired with the AI.

We can separate \cref{reliance-counts} into correct and incorrect explanations. When we only consider cases with correct explanations (\cref{hai-correct} in \cref{hai-appendix}), the non-experts' task accuracy is approximately the same as the AI alone: $48.81\%$ for natural language explanations and $49.33\%$ for example-based explanations. Experts reach CTP in both modalities. When only considering incorrect explanations (\cref{hai-incorrect} in \cref{hai-appendix}), we still see complementary team performance for the experts. However, the non-experts' task accuracy suffers more when shown incorrect explanations. Non-experts' task accuracy for natural language explanations is $42.86\%$ and $36.67\%$ for example-based explanations.

Additionally, we calculate two-sample t-tests to assess whether the trends in \cref{reliance-counts} are significant. The team performance of experts and AI is significantly higher than the team performance of non-experts and AI in both explanation modalities (natural language: p-value $= 0.00$, example-based: p-value $= 0.00$). Furthermore, we also compare the performance for correct and incorrect explanations. Here, we see the same results: experts achieve a significantly higher team performance than non-experts (correct explanations --- natural language: p-value = 0.00, example-based: p-value $= 0.01$; incorrect explanations --- natural language: p-value $= 0.00$, example-based: p-value $= 0.00$).

\subsection{Deception caused by Imperfect XAI}
\label{deception_section}

\begin{figure}[!b]
  \centering
  \includegraphics[trim={0cm 0cm 0cm 0cm},width=0.9\linewidth,clip]{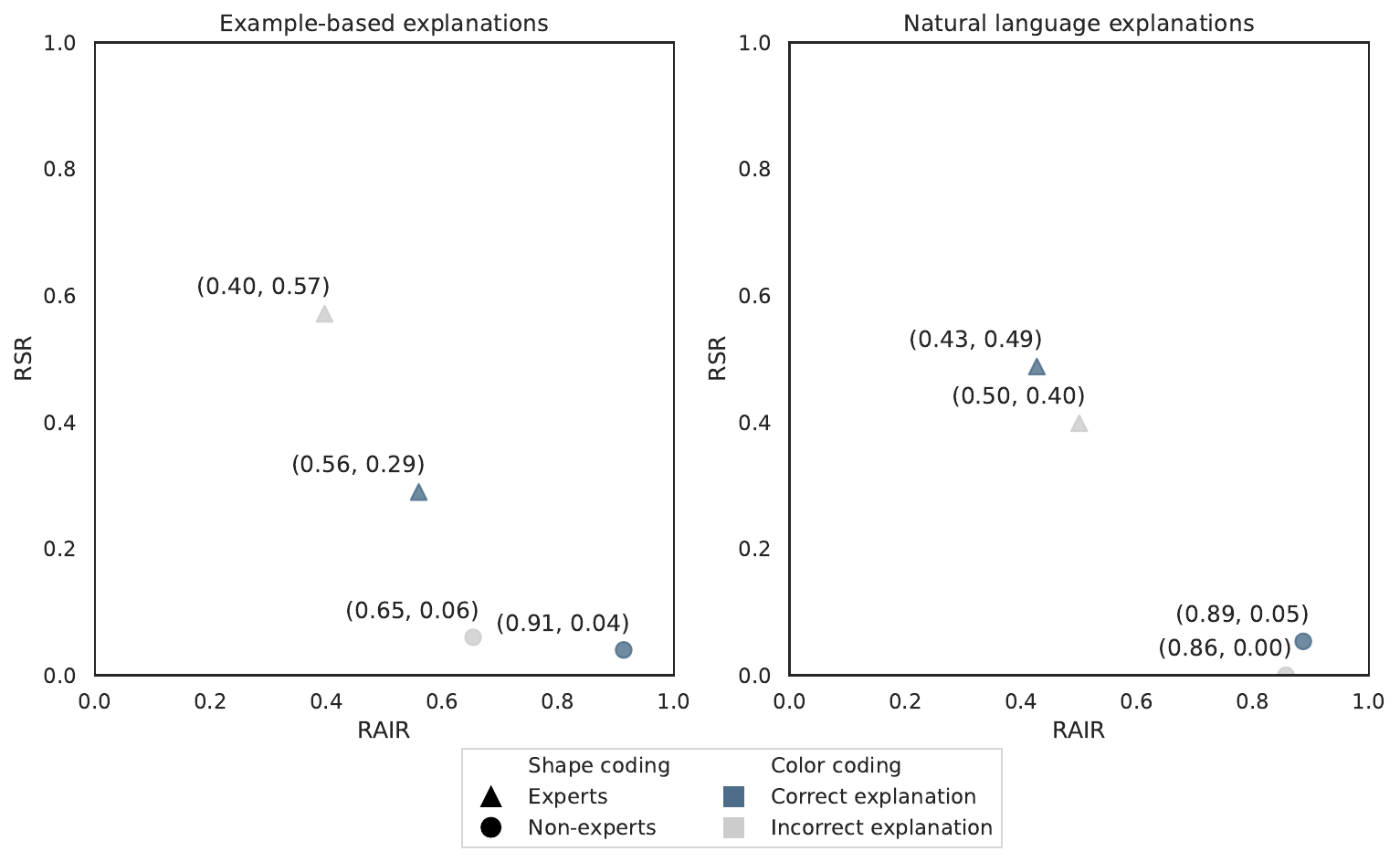}
  \caption{Average observed RAIR (correct AI advice) and RSR (incorrect AI advice) for example-based explanations (on the left side) and natural language explanations (on the right side). We show the average RAIR and RSR for both levels of expertise as well as correct and incorrect explanations.} 
  \label{rsr-rair}
\end{figure}

In \Cref{rsr-rair}, we compare RAIR to RSR for both levels of expertise and the correctness of explanations. We show this comparison for example-based explanations (the graph on the left side of \Cref{rsr-rair}) and natural language explanations (the graph on the right side of \Cref{rsr-rair}). By measuring RAIR and RSR for incorrect and correct explanations separately, we can calculate the deception caused by imperfect XAI (refer to \cref{DAIR_aor_eq} on p. \pageref{DAIR_aor_eq}). We do not visualize assertiveness since we do not see any significant direct or moderation effects. 





The figure shows that experts have a higher RSR than non-experts for both incorrect and correct explanations across both explanation modalities, validating that experts rely more on their own initial decisions when AI advice is given. The most striking result that emerges from the data is that for example-based explanations, we observe that experts have a significantly higher RSR for incorrect explanations (RSR $= 0.57$) than correct explanations (RSR $= 0.29$), resulting in a negative $DoR_{RSR}$ of $-0.28$ (p-value $= 0.00$).
As a result, experts are falsely relying on the AI advice when provided with correct example-based explanations\footnote{Note that correct example-based explanations are consistent in showing three images of the predicted class. Incorrect example-based explanations represent three images that do not correspond to the predicted class of the AI. Moreover, the examples shown are not consistent with the bird species displayed in 90\% of the \textit{correct advice, incorrect explanation} cases and in 40\% of the \textit{incorrect advice, incorrect explanation} cases in our study.}. This means that experts are prone to being misled by correct explanations when the AI advice is incorrect.
However, we do not see this trend for natural language explanations. Here, there is a positive $DoR_{RSR}$ of 0.09, which is not significant (p-value $= 0.41$).
For example-based explanations, the $DoR_{RAIR}$ is positive, meaning that experts rightly follow correct AI advice more often when provided with correct explanations than with incorrect explanations. The data shows a weak, significant positive deception of reliance for example-based explanations ($DoR_{RAIR} = 0.16$, p-value $= 0.10$). Similarly to the RSR cases, for the RAIR cases, the experts are provided with three consistent examples for correct explanations that represent the AI's correctly predicted bird species. The incorrectly provided explanations represent three images that can be inconsistent in the bird species. Thus, experts are deceived by such incorrect explanations even though the AI advice is correct.

Non-experts have, in both modalities, a similar $DoR_{RSR}$ indicating no significant difference in their RSR between correct and incorrect explanations. However, non-experts follow the correct AI advice for correct example-based explanations more often than for incorrect example-based explanations (significant with p-value $= 0.01$). For the latter, the three examples can show inconsistent bird specie(s) that are different from the ground truth of the shown image. Thus, the $DoR_{RAIR}$ for non-experts is at $0.26$. Interestingly, for natural language explanations, the incorrect explanations are not misleading as much ($DoR_{RAIR} = 0.03$, not significant, with a p-value $= 0.68$). This means that non-experts are not misled by incorrect explanations in natural language as much as by visual, example-based explanations. In general, non-experts have a higher RAIR than experts.

Overall, participants have a higher $DoR(RAIR, RSR)$  for example-based explanations (experts: $DoR(RAIR, RSR) = 0.32$; non-experts: $DoR(RAIR, RSR) = 0.26$) than for natural language explanations(experts: $DoR(RAIR, RSR) = 0.11$; non-experts: $DoR(RAIR, RSR) = 0.06$). This means that especially the correctness of example-based explanations has an impact on humans' decision-making behavior.

\subsection{Designing for Imperfect XAI} \label{qual-res}

At the end of the bird identification task, we ask participants: ``\textit{Under what circumstances would you prefer assertive (e.g.,
``definitely'', ``clearly'') versus non-assertive (e.g., ``might be'', ``appears to be'') versus neutral explanations
and why?}''. With imperfect XAI existing in human-AI collaborations, it is necessary not only to understand quantitatively how it impacts decision-makers but also qualitatively. Even though we do not observe the level of assertiveness to have a direct effect or a moderation effect on appropriate reliance, it is still valuable to analyze participants' preferences when it comes to the tone of explanations.

\begin{figure}[t]
  \centering
  \includegraphics[trim={0cm 0cm 0cm 0cm},width=0.9\linewidth,clip]{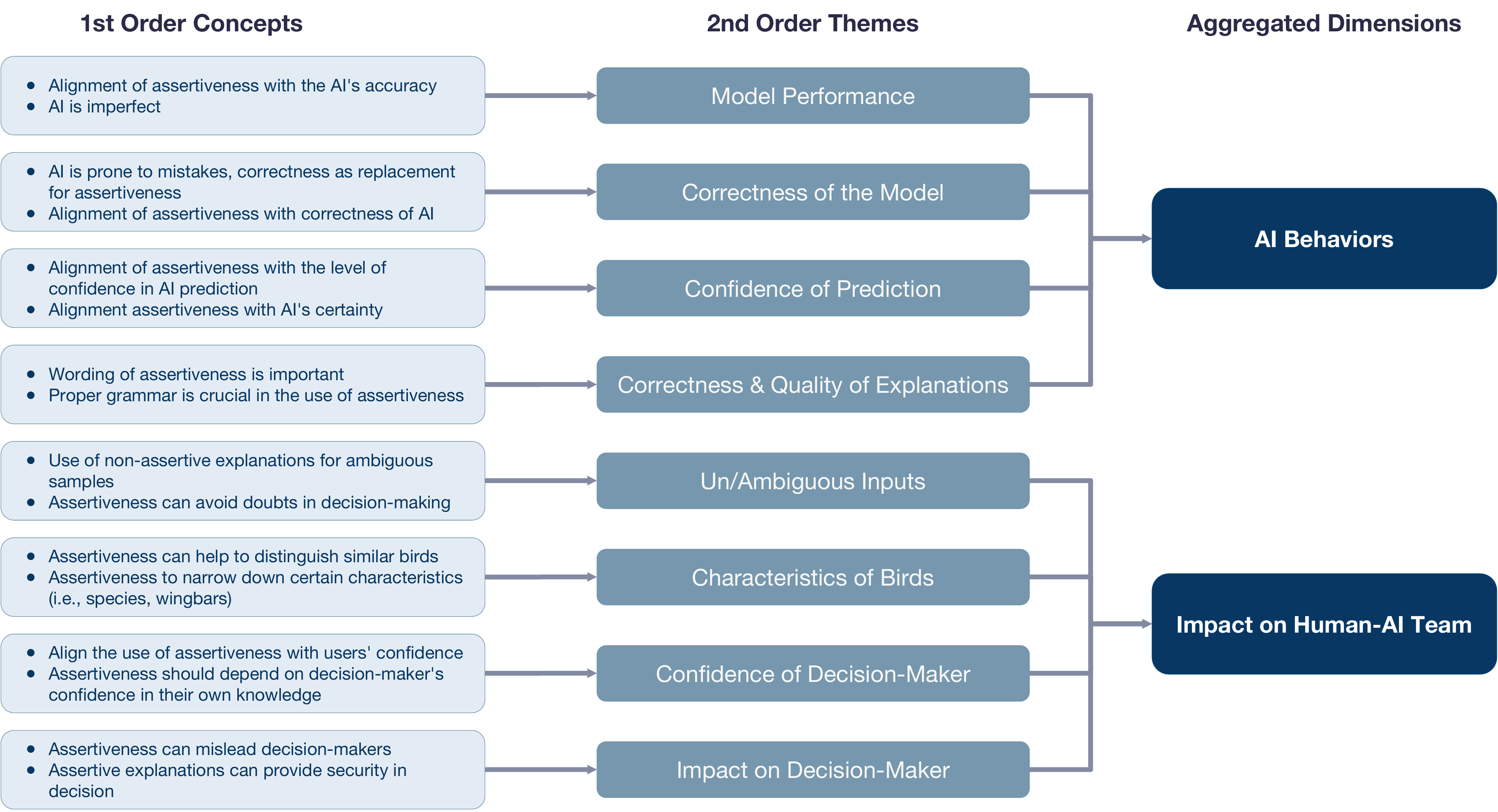}
  \caption{Findings from the qualitative analysis of the survey question: ``Under what circumstances would you prefer assertive (\emph{e.g.},
``definitely'', ``clearly'') versus non-assertive (\emph{e.g.}, ``might be'', ``appears to be'') versus neutral explanations
and why?''. Factors that should be taken into consideration when designing explanations for imperfect XAI systems.
}
  \label{qual-data}
\end{figure}

Through inductive content analysis of participants' responses to this question, we derive two dimensions that researchers and designers should consider when developing and evaluating imperfect XAI in human-AI collaborations: \textit{AI Behaviors} and the \textit{Impact on Human-AI Teams}. Each dimension is made up of four themes that are derived from concepts that emerge in the responses, shown in \cref{qual-data}. We highlight those themes in bold. $25\%$ ($34$ participants) of the responses either do not provide reasoning for their opinion or do not answer the question such that it could be grouped into one of the eight themes we identify. We provide quotes from participants for each theme to structure the aggregated dimensions and shape our insights on designing for imperfect XAI.

\subsubsection{Aggregated Dimension: AI Behaviors}

$54$ out of the $136$ participants answer the survey question with comments relating to the first aggregated dimension: AI Behavior. Participants rationalize that the AI's behavior determines when they prefer assertive, non-assertive, and neutral explanations. Within the AI Behavior dimension, four themes emerge from the participants' comments, such as the model's overall performance, whether the model's prediction is correct or not, the model's confidence in individual predictions, and the correctness/quality of the explanation for a given prediction. 

While only $6$ out of the $136$ participants make comments about the \textbf{model's performance}, it still provides interesting insight that should be considered. Instead of looking at the individual prediction level, these participants focus on the global performance of the model. One participant states that if developers find their model ``\textit{... to be 90\% accurate in your testing, use more definite language, but if it’s not there yet, consider making the tone more neutral and put more responsibility with the end user to interpret the field marks ...}''.

Looking at the individual prediction level, $8$ out of $136$ participants comment on the \textbf{correctness of the AI's prediction}. For example, one participant says that ``\textit{... a non-assertive response would be preferred since the AI selections were incorrect ...}'' while another participant says, ``\textit{I would prefer the assertive language to be accompanied by correct identifications}''.

Considering individual predictions on a more granular level, many participants ($31$ out of $136$) make comments related to \textbf{the confidence of the AI's prediction}. These participants comment on how this factor could be used to determine the tone of the explanation. Specifically, participants, ``\textit{... would prefer the level of assertiveness to depend on the level of confidence of the answer given by the AI}''. One participant expands upon that sentiment by specifying when non-assertive versus assertive tones should be used: ``\textit{I would prefer assertive sentences when the probability of the AI model is very high, while I would prefer non-assertive when the probability is very close to other classes of the model}''. 

$9$ out of $136$ participants make comments related to the \textbf{correctness and quality of the AI explanations}. One participant who sees example-based explanations comments on how some of the examples are incorrect and do not show the correct species. They use this specific situation to rationalize when they would prefer the tone of explanations to be assertive versus non-assertive:  
``\textit{I would prefer assertive explanations if the `similar' photos were actually of the correct species and if the explanation confirmed this. Otherwise, non-assertive explanations are more helpful}''. Another participant who makes a comment related to this theme agrees that assertive language should be used, ``\textit{when all of the reference pictures match up and there are no other similar-looking species}''. 

One participant who sees the natural language explanations comments on the quality and detail of the explanation being a factor to use when determining the tone of the explanation, ``\textit{If the bird description [natural language explanation] is very generic, i.e., brown wings, gray body, or yellow beak (traits that correspond to many birds), I'd rather the AI appear more cautious in its judgment and use non-assertive explanations. However, if the bird has some standout characteristic that the AI correctly identifies [through the natural language explanation], i.e., bright yellow body or red-tipped wings, etc., then assertive explanations seem more convincing}''.

\subsubsection{Aggregated Dimension: Impact on Human-AI Team} 

$48$ out of the $136$ participants answer this survey question with comments related to the human-AI team, such as the confidence and knowledge of the decision-maker, impact on the decision-maker, unambiguous input data, and characteristics of the input data. 

$10$ of $136$ prefer the level of assertiveness to align with their \textbf{own confidence level} and knowledge of the domain. For example, one participant says, ``\textit{I would prefer more assertive explanations when I don't feel very sure about my choice}''. Another participant adds that they would prefer assertive explanations if they ``\textit{... didn't know anything about the topic in question ...}''. However, one participant says, ``\textit{I would prefer neutral explanations if I'm unsure of the species and assertive if I'm confident in my identification}''.



$15$ out of the $136$ make comments related to the \textbf{impacts on the decision-maker}. Some comments consist of concerns related to being misled and over- or under-relying on the AI, while other comments motivate the benefits of having assertive explanations. For example, one participant states that ``\textit{Assertive words create more security while changing your opinion or trying to gain knowledge}''. Another participant who shares the same sentiment said, ``\textit{I would prefer assertive explanations because it would make me feel more secure about the answer}''.
However, some participants do not share the same sentiment about assertive explanations and pointed out the potential consequences of them: ``\textit{Assertive AI explanations were given for incorrect identifications, which would mislead users}''. Given that potential to be misled, another participant rationalizes they ``\textit{... would prefer non-assertive explanations because I [they] do not fully trust AI with bird ID just yet}''.

$17$ out of the $136$ participants rationalize that the assertiveness of explanations should be based on how \textbf{ambiguous the input} is for a given prediction. For example, one participant brings up the quality of the input image and the difficulty of the bird ID as a way to determine whether explanations should be assertive or not: ``\textit{When it comes to less distinctive IDs like most sparrows, or harder to ID circumstances like winter or females or juveniles, or situations with weird lighting or harder angles it makes sense to use non-assertive.}''. On a similar line of thought, another participant says, ``\textit{I would prefer assertive explanations for birds that have distinctive traits over similar bird species, and non-assertive or neutral explanations for birds that are similar with characteristics that are more difficult to tell apart}''.


$6$ out of the $136$ participants commented on how the use of assertive explanations helped them realize various \textbf{characteristics of the birds}. For example, one participant values the assertive tone because it is ``\textit{... helpful for pointing out distinctive features that would help ID the bird.}''. They also think that ``\textit{... the non-assertive language was helpful for species that share similar characteristics (aka clear, non-streaked breast) with other species that share that characteristic}''.
\section{Discussion}
\label{discussion_section}


We investigate how imperfect XAI impacts humans' decision-making when collaborating with an AI. More precisely, we assess how imperfect explanations affect humans' reliance behavior and investigate the effects on the human-AI team performance. To answer RQ \ref{rq1} and RQ \ref{rq2}, we assess the validity of our research model for two different types of explanations: natural language explanations and example-based explanations. Previous research emphasizes the need to consider imperfect AI when designing for human-AI collaboration~\cite{kocielnik2019will}. With recent research looking into how humans and AI can achieve complementary team performance~\cite{bansal2021does},  \citet{schemmer2023appropriate} conceptualize the role of appropriate reliance in human-AI collaboration. We extend \citet{schemmer2023appropriate}'s framework by adding another dimension: XAI advice. Given that an explanation can be incorrect even if the AI advice is correct, it is crucial to understand the impact of incorrect XAI advice on decision-making. Furthermore, it is necessary to understand the impact of imperfect XAI for different types of explanations. Below, we discuss how our contributions are situated in current literature and the implications for CSCW. 

In our study, we observe a significant moderation of humans' level of expertise on the effect of explanations' correctness on RAIR for both explanation modalities. However, we do not see this moderation for RSR. When humans are being provided wrong AI advice, their level of expertise does not moderate the impact of imperfect explanations on humans' RSR. We identify a direct effect of the level of expertise on RSR in both explanation modalities. Additionally, the correctness of explanations impacts RSR negatively for example-based explanations. Overall, our work synthesizes how humans' level of expertise impacts their reliance on AI when provided with imperfect explanations. Non-experts rely more on AI than experts, whereas experts rely more on their initial decisions. Especially for example-based explanations, imperfect XAI deceives experts' self-reliance and experts'/non-experts' relative AI reliance, fostering inappropriate reliance on the AI. Thus, this study sets a starting point to investigate the effect of imperfect XAI for different explanation modalities.

\textbf{Our findings show that imperfect explanations impact human-AI decision-making}. We observe that experts reach complementary team performance when imperfect explanations are provided. This holds true for natural language and example-based explanations. While non-experts do not reach complementary team performance, our analyses reveal that their performance can be improved to be similar to that of the AI performance. 
Moreover, there is a difference between reaching complementary team performance when the correctness, or fidelity, of the explanation changes (see \Cref{hai-correct} and \Cref{hai-incorrect} in \Cref{hai-appendix}). Previous research discusses the impact of explanations' fidelity on humans' reliance on AI and hypothesizes that fidelity has a positive impact on humans' reliance behavior on AI \cite{hemmer2021human}. With our results, we confirm this hypothesis. Furthermore, ~\citet{papenmeier2019model} observe that low-fidelity explanations (or incorrect explanations) impact user trust in AI when the global model performance is around $75\%$ accurate, which helps validate our findings.
We also observe that the lack of expertise among non-experts impacts their task performance when shown incorrect explanations regardless of the AI advice being correct (\cref{hai-incorrect} in \cref{hai-appendix}). Similar to our findings, \citet{nourani2020role} observe that non-experts tend to over-rely on AI advice, attributing this to their inability to identify when the AI is incorrect because of their lack of expertise. These findings contribute to a more integrated understanding of the impact of human-AI decision-making on different user groups in the presence of imperfect XAI. For example, this can inform managers on how to assign tasks to humans with different levels of expertise and provide them with explanations in different modalities. \kmedit{It could also lead to organizations modifying their human-AI collaboration workflows. From informal conversations with the product team of an AI decision-support tool\footnote{WildMe.org} for biologists and conservationists to classify species and identify individuals from camera trap imagery, we learned that organizations using their tool have modified their workflow to incorporate ``checks-and-balances''. For example, intro-level biologists will collaborate with the AI to match individuals and then request a review of their ``human-AI team'' decision from a higher-up. In this unique human-human-AI collaboration scenario, the expert biologist could potentially correct situations when an intro-level biologist over-relies on AI advice because of an incorrect explanation.}

\textbf{Visual, example-based explanations are more deceptive than natural language explanations.} 
To account for the impact of imperfect XAI on humans' appropriate reliance, we establish a novel metric \textbf{DoR}, (Deception of Reliance), to measure the difference in RAIR and RSR for correct and incorrect explanations. Our results indicate that people are more deceived by example-based explanations than by natural language explanations. In terms of RAIR (note that in RAIR cases, the AI advice is correct), experts and non-experts are deceived by incorrect explanations. In terms of RSR (note that in RSR cases, the AI advice is incorrect), experts are deceived by correct explanations. This is an interesting observation that may be explained by the consistency of shown visual examples. For \textit{correct advice, incorrect explanations} cases, the XAI is providing three visual examples that show a different bird species than the bird species on the image to be classified (see \Cref{explanationformat}). Moreover, these three visual examples can belong to different bird species since the XAI is choosing the top three most similar bird images (in our study, this is in 90\% of all \textit{correct advice, incorrect explanation} cases). This inconsistency in example-based explanations might deceive experts and non-experts to not rely on the AI anymore when they identify visual differences in the images provided as explanations, disregarding the correct AI advice. We discover the same behavior for experts for \textit{incorrect advice, correct explanation} cases. In those cases, the explanations consist of three images of the same bird species as the AI predicted. The incorrect explanations consist of three images that can be inconsistent in the bird species shown (in our study, this in 40\% of all \textit{incorrect advice, incorrect explanation} cases). Thus, this inconsistency in examples might deceive experts into no longer relying on themselves anymore when they identify three consistent examples shown, disregarding the incorrect AI advice. Hence, the DoR of experts and non-experts is positive for RAIR cases as they are deceived by incorrect explanations, while experts additionally have a negative DoR and are deceived by correct explanations. Note that the overall RSR for experts is still higher than non-experts' RSR; the impact on deception caused by imperfect XAI is higher.
As participants mentioned in the survey, it is more convincing that there is less uncertainty in the AI advice when three images that are similar to each other are shown than when three different images are shown. This corroborates our findings. This trend is not present in natural language explanations. 

\textbf{The language tone of explanations does not impact humans' decision-making behavior. } Calisto et al. conclude that the level of expertise influences whether the framing of the explanation should be assertive or non-assertive~\cite{calisto2023assertiveness}. Based on their observations, they specifically suggest that natural language explanations should be designed such that the tone of the explanation is appropriate for the end user's level of expertise.
Despite the numerous previous studies finding that the framing of the explanations has a significant impact on human-AI collaboration~\cite{kim2020effect,calisto2023assertiveness,kim2023communicating}, our findings do not show an impact on appropriate reliance. Quantitatively, we do not find a direct effect of assertiveness on appropriate reliance. Qualitatively, we observe that the tone of the explanation should depend on certain situations, such as the AI's behaviors, instead of the human's level of domain expertise. We observe that $31\%$ of the participants prefer assertiveness to align with the confidence of the model's prediction, while only $10\%$ of participants prefer assertiveness to align with their confidence in the domain. This finding could be attributed to the fact that participants are collaborating with imperfect AI and are able to acknowledge that the AI advice was occasionally incorrect. However, given these qualitative suggestions \kmedit{and the potential for natural language explanations to present irrelevant or incorrect information~\cite{sovrano2023toward}}, we encourage future work to explore various ways to alter the tone of an explanation based on the themes we identified in \cref{qual-data}. 

\textbf{Our findings can guide researchers and practitioners on how to assess and design for imperfect XAI in human-AI collaborations.}
Regardless of the explanation modality, it is important to understand how humans interact with imperfect XAI. Visual explanations, such as example-based explanations and saliency maps, have been shown in the past to be of high educational value to the end-user (\emph{e.g.}, \cite{kim2023help,mac2018teaching}), making it even more important to understand how to design for and mitigate imperfect XAI. This need is intensified with the role AI takes in organizational learning \cite{spitzer2023ml}. Especially in the workplace, AI can facilitate knowledge transfer and support organizations in retaining and distributing expert knowledge~\cite{spitzer2022training, jarrahi2023artificial, wilkens2020artificial}. Similarly, it is also crucial to understand how imperfect XAI affects the learning of novices through AI-based learning systems \cite{spitzer2023ml} or through collaboration with AI \cite{schemmer2023towards}.
Our findings can guide knowledge managers within organizations on how to make use of explanations for employees with different levels of domain knowledge. More precisely, knowledge managers should be aware of the impact of exposing humans with different levels of expertise to imperfect XAI.
In addition, our findings contribute to a more integrated understanding of the impact that incorrect explanations can have on human-AI decision-making and inform different stakeholders in organizations. As non-experts are more affected by imperfect XAI, their performance drops more than experts' performance when incorrect explanations are provided in comparison to correct explanations. With this finding, knowledge managers can adjust their knowledge retention activities when training new employees; designers can adjust the development of human-AI collaboration systems to successfully facilitate explanations in decision-making and support humans in their work setting. Thus, we encourage practitioners designing human-AI collaboration systems to apply our findings to structure their design approach. This can aid organizations in laying out the strategic direction of human resource development by matching the use of explanations to humans' prior knowledge. Overall, these findings shed light on the ongoing discussion in CSCW on how to make use of explanations within work settings.

\section{Limitations \& Future Work}
\label{limitations}

We elaborate on various limitations of our study, how they could impact the interpretation of our results, and identify opportunities for future work.

\textbf{Lack of Information to Properly Identify Birds.} Expert birders usually rely on more information than just the visual characteristics of a bird when determining the bird species, especially for ambiguous cases. For example, the location and habitat in which the bird was spotted can be imperative to determine the exact bird species within a family. It's unclear to what extent the lack of this information influences our results. Future technical work could consider using this information to help build more transparent bird classification models.

\textbf{Correctness of Explanations versus Explanation Fidelity.} Throughout our study, we consider explanations to either be incorrect or correct. However, as we mention, some explanations that we classify as incorrect can contain evidence that is correct, making it difficult to only have a binary categorization for the correctness of explanations. While we use a binary scale for our analyses, explanation correctness, or fidelity, can be quantitatively measured on a continuous scale and categorized as low fidelity and high fidelity~\cite{papenmeier2019model}. We encourage future work to explore explanation fidelity using multiple categories instead of two to understand the differences between low- and medium-fidelity explanations when it comes to task performance and appropriate reliance. This will provide insight into the impact that noisy explanations have on decision-making, such as when an explanation reveals some information that is aligned with the ground truth class and some information that is aligned with the predicted class.

\textbf{Simplistic Natural Language Explanations.} 
The natural language explanations are very limited and simplistic, although this is a fault of the model that we are using to generate those natural language explanations~\cite{hendricks2021generating}. Compared to a field guide such as All About Birds~\cite{birdsguide}, these natural language explanations could be correct for several species given their lack of detail and their short length. This is potentially a side effect of the text descriptions in the CUB-200-2011 dataset being sourced from crowd workers instead of experts. We encourage technical researchers who are developing explanation methods that explain image classifications using natural language to pay careful attention to the text descriptions used for training the language model. For example, researchers could consider using text data from a field guide as their training data. We also encourage CSCW researchers to conduct more studies on how the explanation's length and level of detail impact experts' and non-experts' appropriate reliance. 

\textbf{Different Explanations Convey Different Information.} 
While we do not intend to directly compare the two explanation modalities throughout our analyses, they are discussed in terms of similarities and differences throughout the article. Our main intention is to investigate how our research model holds across different modalities of explanations. While several previous studies compare multiple different types of explanation modalities qualitatively and quantitatively (\emph{e.g.}, ~\cite{kim2023help,chen2023understanding,kim2023should,du2022role,szymanski2021visual}), we encourage readers to avoid directly comparing the two explanations because they present different information. Previous research reveals that different explanation techniques can result in disagreements for the same dataset \cite{roy2022don}. For example, the natural language explanations from \citet{hendricks2021generating} are feature-based, providing descriptions of features present in the image~\cite{hendricks2021generating}. However, the example-based explanations present three similar images, which is very different information from the natural language description of features. \psedit{On top of that, the incorrectness in both explanation modalities is represented in different ways. While there are factual errors in natural language explanations that previous research addresses (e.g., hallucination effects of natural language models~\cite{sovrano2023toward}), there are logical errors (e.g., inconsistencies) within example-based explanations. This opens avenues for future research to investigate how different human cognitive abilities (i.e., cognitive styles) impact the perception of these imperfect explanations in AI-assisted decision-making scenarios.}

\textbf{Visualizing Assertiveness for Example-Based Explanations.} While there is previous research on how to visualize the confidence of a prediction for image classification~\cite{tejeda2023displaying}, to the best of our knowledge, there is no work visualizing the assertiveness of example-based explanations. Given this, our example-based explanations use natural language to convey assertiveness. Visualizing the assertiveness for each example in an example-based explanation would provide the decision-maker with another queue about whether they should rely on the model. 

\textbf{AI Expertise versus Domain Expertise.} Our analyses are based on the participant's expertise in bird species identification. We do not ask participants about their knowledge of AI. Thus, we do not analyze their expertise related to AI. It is unclear to what extent the participant's expertise with AI would impact the effect of imperfect XAI on appropriate reliance and task accuracy for bird species classification. Future work should consider looking at the effect that AI knowledge combined with domain expertise has on appropriate reliance, taking into account an imperfect XAI.
\section{Conclusion}
\label{conclusion_section}

This article sets out a research model to investigate the effect of imperfect XAI on human-AI decision-making. Thus far, human-computer interaction and CSCW literature fail to thoroughly scrutinize how explanations' correctness affects humans' decision-making and their reliance behavior on AI. Hence, through a human study with 136 participants, we empirically analyze humans' decision-making and specifically assess whether their level of expertise and explanations' assertiveness moderate the effect of imperfect XAI on appropriate reliance. Furthermore, we explore to what extent incorrect explanations deceive decision-makers' reliance on AI. With our findings, we make several contributions: First, we propose a research model to investigate the moderation of assertiveness and humans' level of expertise on imperfect XAI in decision-making tasks. We thereby extend the existing conceptualization of appropriate reliance by a new dimension of XAI advice. Second, through an empirical study, we reveal that imperfect explanations and participants' level of expertise affect human-AI decision-making for two different explanation modalities. In addition, we show the effect on complementary team performance and provide guidance for future studies on how to investigate imperfect XAI in the context of human-AI decision-making. Third, we propose a novel metric called Deception of Reliance (DoR), which allows us to measure the impact of incorrect explanations on decision-makers' reliance. Our results inform designers of human-AI collaboration systems and provide guidelines for their development. Fourth, we reveal which role the language tone in explanations plays and outline important dimensions that should be considered when designing for XAI advice.

Overall, with this work, we reveal the impact of imperfect XAI on human-AI decision-making by taking into account humans' level of expertise and explanations' assertiveness. Extensive and rigorous research is needed to fully understand and exploit imperfect XAI in decision-making. We invite researchers to take part in this debate and hope to inspire scientists to actively participate in this endeavor.

\begin{acks}
We would like to thank Youwei Jiang for helping with the development of portions of the user interface used for the study. We would like to thank Hao-Fei Cheng and Haiyi Zhu for providing feedback on the initial study design. We also thank Max Schemmer and other lab members of the KSRI for discussing our study design and findings with us. 
We would like to thank Nari Johnson, Hayden Stec, Adel Sharif, Donny Bertucci, Alex Cabrera, Will Epperson, Venkat Sivaraman, and other members of the DIG Lab at CMU for participating in our initial pilot studies. We would like to thank all of the experienced birders who provided us with feedback on our study design and discussed the limitations of our study with us. Lastly, we would like to thank all of our connections that helped us recruit birders. Research reported in this publication was supported by the National Heart, Lung, and Blood Institute of the National Institutes of Health under award number R01HL164906. ChatGPT was utilized to generate LaTeX code for some of the tables in this paper.
\end{acks}

\bibliographystyle{ACM-Reference-Format}
\bibliography{references}

\appendix
\newpage
\section{Appendix}

\subsection{Appropriate Reliance with Imperfect XAI}
\label{formulas}

\begin{table}[H]

\caption{Overview of the newly introduced metrics when considering imperfect explanations.}
    \begin{threeparttable}
\begin{tabular}{P{2cm} m{11cm} }
 \hline
$CSR_{IC}$ & Correct self-reliance for the case where the AI gives incorrect advice and a correct explanation is one when the initial human decision is correct and the final decision is correct. \\ \hline 
$OR_{IC}$ & Over-reliance for the case where the AI gives incorrect advice and a correct explanation is one when the initial human decision is correct and the final decision is correct. \\ \hline 
$CSR_{II}$ & Correct self-reliance for the case where the AI gives incorrect advice and an incorrect explanation is one when the initial human decision is correct and the final decision is correct. \\ \hline 
$OR_{II}$ & Over-reliance for the case where the AI gives incorrect advice and an incorrect explanation is one when the initial human decision is correct and the final decision is correct. \\ \hline 
$CAIR_{CC}$ & Correct AI reliance for the case where the AI gives correct advice and a correct explanation is one when the initial human decision is incorrect and the final decision is correct. \\ \hline 
$UR_{CC}$ & Under-reliance for the case where the AI gives correct advice and a correct explanation is one when the initial human decision is incorrect and the final decision is correct. \\ \hline 
$CAIR_{CI}$ & Correct AI reliance for the case where the AI gives correct advice and an incorrect explanation is one when the initial human decision is incorrect and the final decision is correct. \\ \hline 
$UR_{CI}$ & Under-reliance for the case where the AI gives correct advice and an incorrect explanation is one when the initial human decision is incorrect and the final decision is correct. \\ \hline 
\end{tabular}
    \begin{tablenotes}
        \item[1] A correct AI explanation correpsonds with the AI's advice, no matter if the advice is correct or incorrect. For a classification task this means the following: If the AI gives incorrect advice and the explanation is correct, the explanation aligns with the incorrectly predicted class.
    \end{tablenotes}
    \end{threeparttable}

\label{mod_anal_nle_rair}

\end{table}

Following the newly introduced dimension, the calculation for RAIR and RSR are adjusted to the following:
\begin{equation}
    RSR\ (Relative\ Self-Reliance) = \frac{\sum_{i=0}^{N}{(CSR_{IC, i} + CSR_{II, i})}}{\sum_{i=0}^{N}{IA_i}}
\end{equation}
\begin{equation}
    RAIR\ (Relative\ AI\ Reliance) = \frac{\sum_{i=0}^{N}{(CAIR_{CC, i} + CAIR_{CI, i})}}{\sum_{i=0}^{N}{CA_i}}
\end{equation}
\newpage
\subsection{Bird Identification Test}
\label{bird-test-details}

The bird identification test consists of images of six images: three ``easy'' common bird species and three ``hard'' bird species. The three ``easy'' common bird species were selected with the intention that most beginning birders would be familiar with them. For the ``easy'' common bird species, participants have to identify a \textit{Downy Woodpecker}, a \textit{Herring Gull}, and a \textit{Ruby-Throated Hummingbird}.  

For the ``hard'' bird species, participants have to identify a \textit{female Hooded Warbler}, a \textit{Blue-headed Vireo}, and a \textit{Chestnut-sided Warbler}. The \textit{female hooded warbler} is chosen because it looks significantly different than a male Hooded Warbler and requires a higher level of expertise to be able to correctly identify that. The \textit{Blue-headed Vireo} is chosen because it visually looks very similar to the \textit{Philadelphia Vireo}, again requiring a higher level of expertise to correctly identify that. Lastly, the \textit{Chestnut-sided Warbler} is chosen because there are several different species in the Warbler family, and they are easy for non-experts to mix up.

Below are figures showing the performance on the bird test based on the experts and non-experts grouping we do. 

\begin{figure}[H]
  \centering
  \includegraphics[trim={0cm 17cm 13cm 0cm},width=\linewidth,clip]{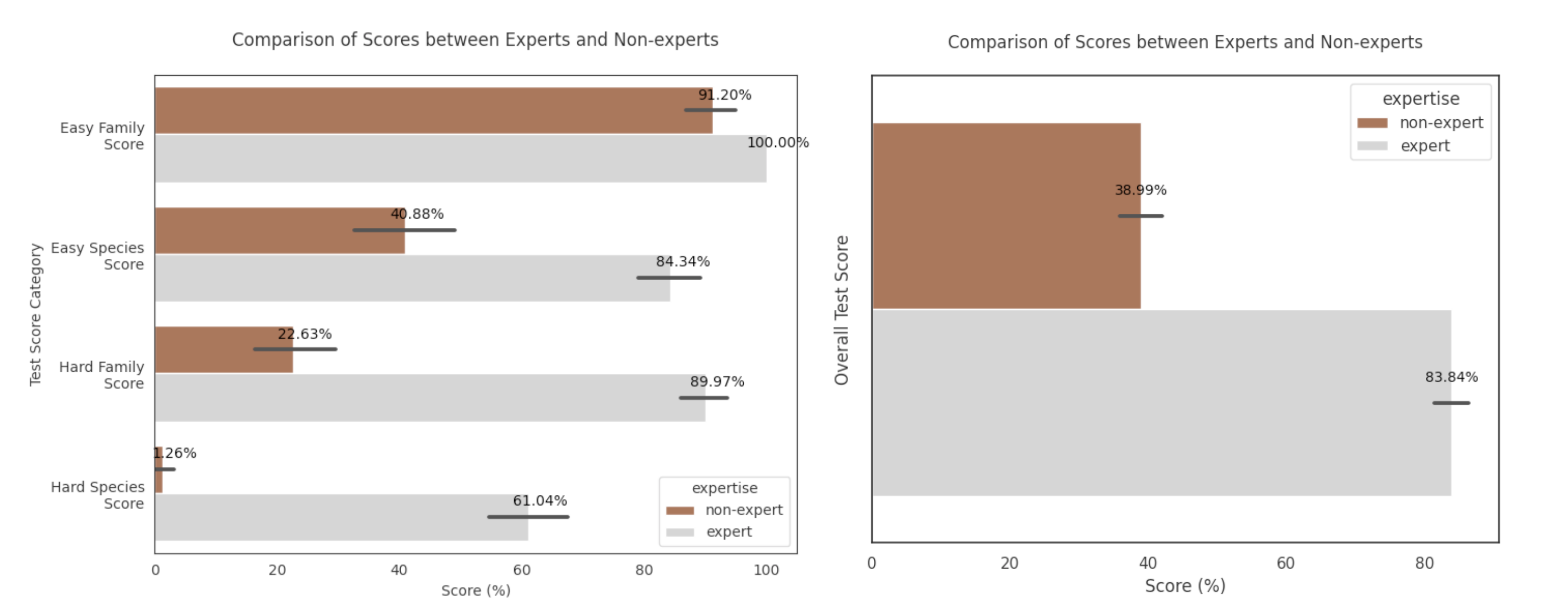}
  \caption{The left half of this figure shows how participants (experts and non-experts) perform on average for the easy birds and three hard birds. We calculate their family accuracy as well as species accuracy. The right half of this figure shows the average overall score by combining the four scores from the left. }
  \label{test-scores}
\end{figure}



\newpage
\subsection{Moderation Analyses}
\label{mod_appendix}
\begin{table}[htbp!]

\caption{Moderation analysis of correctness of natural language explanations on RAIR with the level of expertise and assertiveness as moderators (Since the level of expertise is a three-level categorical moderator, it is split up into Z1 --- non-assertive explanations in relation to all other values --- and Z2 --- assertive explanations in relation to all other values.}
    \begin{threeparttable}
\begin{tabular}{m{1.5cm} R{1.2cm} R{1.2cm} R{1.2cm} R{1.2cm} R{1.2cm} R{1.2cm}}
\\ \hline
 & coeff & ce & Z & p & LLCI & UCLI  \\ \hline \hline
const   & 1.26 & .34 & 3.67   & .00  & .59  & 1.94     \\\hline
corr  & .57 & .53 & 1.07 & .29  & -.48  & 1.62   \\\hline
exp  & -2.12 & .33 & -6.50 & .00  & -2.77  & -1.48  \\\hline
Z1 & -.46 & .39 & -1.16 & .24  & -1.23  & .31  \\\hline   
Z2 & .00 & .39  & .00 & 1.00  & -.76  & .76  \\\hline 
exp x corr & -1.00 & .51 & -1.96  & .05  & -1.99  & .00  \\\hline 
Z1 x corr & .46 & .59  & .77 & .44  & -.70  & 1.61  \\\hline 
Z2 x corr & .19 & .58 & .33  & .74  & -.95  & 1.34 \\\hline 
\end{tabular}
    \begin{tablenotes}
        \item[1] \textit{corr} --- \textit{correctness}; \textit{exp} --- \textit{level of expertise}
    \end{tablenotes}
    \end{threeparttable}
\label{mod_anal_nle_rair}

\end{table}

\begin{table}[htbp!]

\caption{Moderation analysis of correctness of example-based explanations on RAIR with level of expertise and assertiveness as moderators (Since level of expertise is a three-level categorical moderator, it is split up into Z1 --- non-assertive explanations in relation to all other values --- and Z2 --- assertive explanations in relation to all other values.}
    \begin{threeparttable}
\begin{tabular}{m{1.5cm} R{1.2cm} R{1.2cm} R{1.2cm} R{1.2cm} R{1.2cm} R{1.2cm}}
\\ \hline
 & coeff & ce & Z & p & LLCI & UCLI  \\ \hline \hline
const   & .43 & .32 & 1.35   & .17  & -.19  & 1.05     \\\hline
corr  & 1.02   & .49  & 2.08  & .04 & .06 & 1.99   \\\hline
exp  & -1.25  & .31  & -4.09  & .00 & -1.85 & -.65  \\\hline
Z1 & .26 & .36  & .73 & .47 & -.45  & .98  \\\hline   
Z2 & .00 & .37  & .00  & 1.00 & -.72 & .72  \\\hline 
exp x corr & -1.04 & .47 & -2.21  & .03  & -1.95  & -.12  \\\hline 
Z1 x corr & -.03  & .54  & -.06 & .95 & -1.08  & 1.02  \\\hline 
Z2 x corr & -.16  & .54  & -.29  & .77 & -1.22 & .90 \\\hline 
\end{tabular}

    \begin{tablenotes}
        \item[1] \textit{corr} --- \textit{correctness}; \textit{exp} --- \textit{level of expertise}
    \end{tablenotes}
    \end{threeparttable}

\label{mod_anal_ex_rair}

\end{table}

\begin{table}[htbp!]

\caption{Moderation analysis of correctness of natural language explanations on RSR with level of expertise and assertiveness as moderators (Since level of expertise is a three-level categorical moderator, it is split up into Z1 --- non-assertive explanations in relation to all other values --- and Z2 --- assertive explanations in relation to all other values.}

    \begin{threeparttable}
\begin{tabular}{m{1.5cm} R{1.2cm} R{1.2cm} R{1.2cm} R{1.2cm} R{1.2cm} R{1.2cm}}
\\ \hline
 & coeff & ce & Z & p & LLCI & UCLI  \\ \hline \hline
const   & -17.16   & 592.16  & -.03  & .98 & -1177.77 & 1143.45     \\\hline
corr  & 13.25   & 592.16  & .02  & .98 & -1147.37 & 1173.86   \\\hline
exp  & 15.89  & 592.16  & .03  & .98 & -1144.72 & 1176.50  \\\hline
Z1 & .14 & .52  & -.26  & .79 & -.89 & 1.16  \\\hline   
Z2 & -.31 & .56 & -.56  & .58 & -1.41 & .79  \\\hline 
exp x corr & -13.33  & 592.16  & -.02  & .98 & -1173.94 & 1147.29  \\\hline 
Z1 x corr & -.28  & .75  & -.37  & .71 & -1.75 & 1.19  \\\hline 
Z2 x corr & .90  & .74  & 1.20  & .23 & -.56 & 2.36 \\\hline 
\end{tabular}

    \begin{tablenotes}
        \item[1] \textit{corr} --- \textit{correctness}; \textit{exp} --- \textit{level of expertise}
    \end{tablenotes}
    \end{threeparttable}

\label{mod_anal_nle_rsr}

\end{table}

\begin{table}[htbp!]

\caption{Moderation analysis of correctness of example-based explanations on RSR with the level of expertise and assertiveness as moderators (Since level of expertise is a three-level categorical moderator, it is split up into Z1 --- non-assertive explanations in relation to all other values --- and Z2 --- assertive explanations in relation to all other values.}

    \begin{threeparttable}
\begin{tabular}{m{1.5cm} R{1.2cm} R{1.2cm} R{1.2cm} R{1.2cm} R{1.2cm} R{1.2cm}}
\\ \hline
 & coeff & ce & Z & p & LLCI & UCLI  \\ \hline \hline
const   & -3.60 & .76 & -4.74   & .00  & -5.09  & -2.11     \\\hline
corr  & -.44   & 1.29  & -.34  & .74 & -2.97 & 2.10   \\\hline
exp  & 3.25 & .74  & 4.39  & .00 & 1.80 & 4.70  \\\hline
Z1 & -.09 & .43  & -.22  & .83 & -.94 & .75  \\\hline   
Z2 & .09 & .43 & .21  & .83 & -.75 & .93  \\\hline 
exp x corr & -.73  & 1.28  & -.57  & .57 & -3.23 & 1.77  \\\hline 
Z1 x corr & -.45  & .75  & -.60  & .55 & -1.92 & 1.02  \\\hline 
Z2 x corr & -.43  & .72  & -.59  & .55 & -1.85 & .99 \\\hline 
\end{tabular}
    \begin{tablenotes}
        \item[1] \textit{corr} --- \textit{correctness}; \textit{exp} --- \textit{level of expertise}
    \end{tablenotes}
    \end{threeparttable}
\label{mod_anal_ex_rsr}

\end{table}

\newpage

\subsection{Human-AI Team Performance}
\label{hai-appendix}

\begin{figure}[H]
  \centering
  \includegraphics[width=\linewidth,clip]{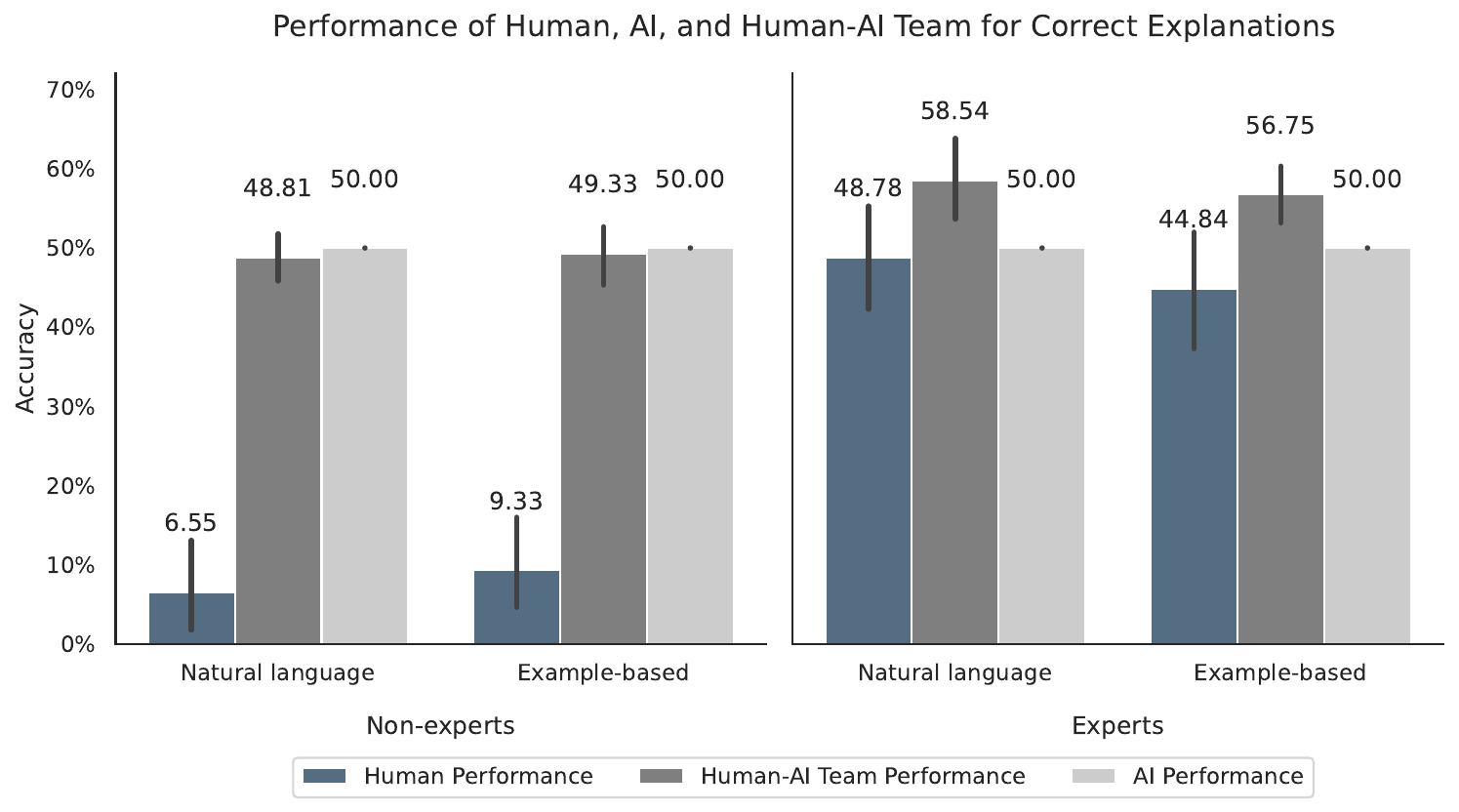}
  \caption{Performance of the human, AI, and human-AI team specifically for correct explanations. This represents 6 birds from the 12 that participants saw.}
  \label{hai-correct}
\end{figure}

\begin{figure}[htbp]
  \centering
  \includegraphics[width=0.9\linewidth,clip]{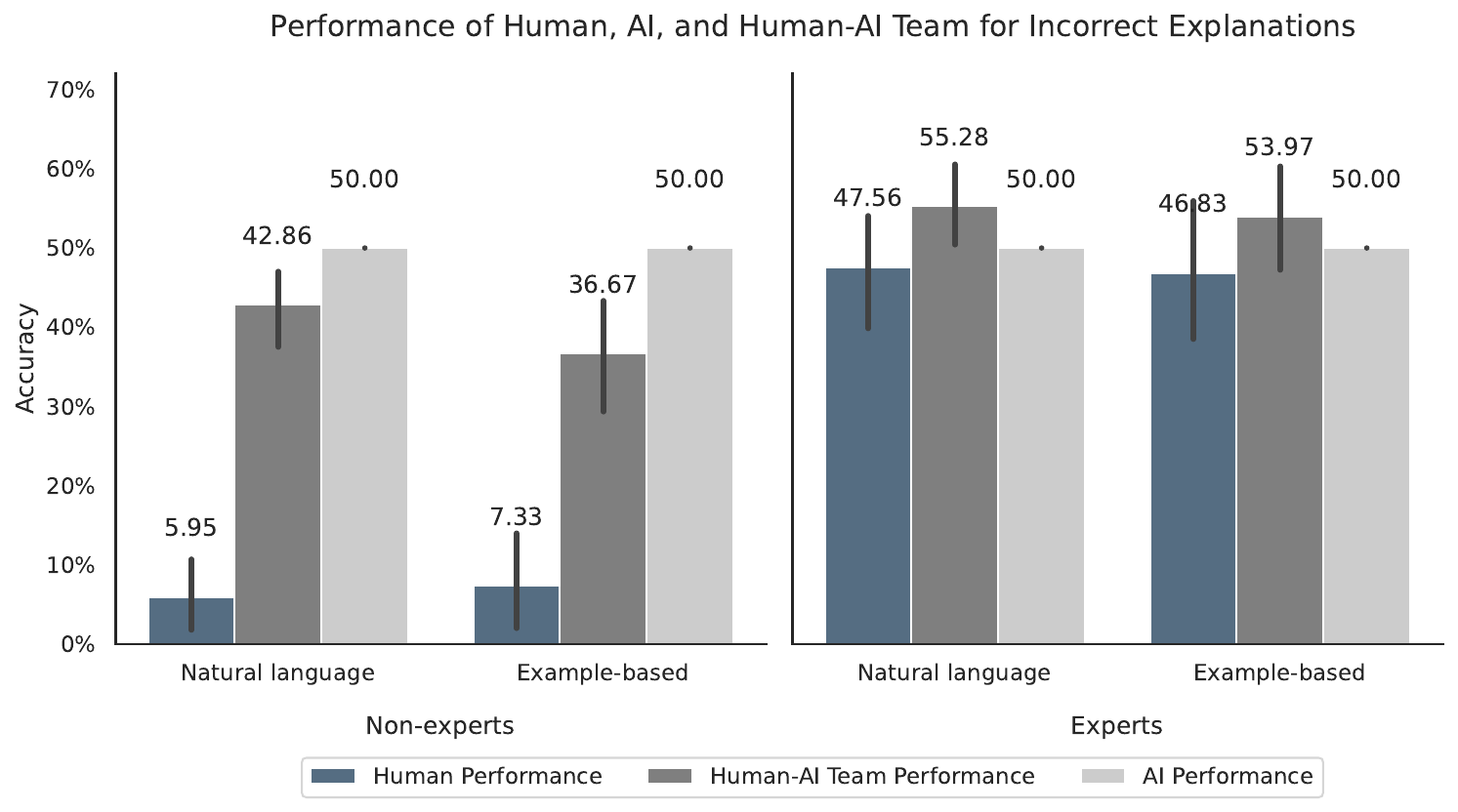}
  \caption{Performance of the human, AI, and human-AI team specifically for incorrect explanations. This represents 6 birds from the 12 that participants saw.}
  \label{hai-incorrect}
\end{figure}

\end{document}